\documentclass[twoside,english]{elsarticle}
\usepackage[utf8]{inputenc}
\usepackage[T1]{fontenc}
\usepackage{geometry}
\usepackage{dsfont}
\usepackage{color} 
\usepackage{amsmath}
\usepackage{amssymb}
\usepackage{graphicx}
\usepackage{cancel}

\makeatletter

\def\ps@pprintTitle{%
	\let\@oddhead\@empty
	\let\@evenhead\@empty
	\def\@oddfoot{}%
	\let\@evenfoot\@oddfoot}

\usepackage{babel}
\usepackage{hyperref}
\usepackage[normalem]{ulem} 

\usepackage{array}
\usepackage{booktabs}
\usepackage{multirow}
\usepackage{rotating}

\makeatother

\newcommand{\chg}[2][black]{{\color{#1}#2}}

\begin{document}

\begin{frontmatter}{}

\title{\chg{Mechanical Feedback in Regulating the Size of Growing Multicellular Spheroids}}

\author[LiPhy]{A.~Erlich\corref{cor1}}

\ead{alexander.erlich@univ-grenoble-alpes.fr}

\author[LiPhy]{P.~Recho\corref{cor1}}

\ead{pierre.recho@univ-grenoble-alpes.fr}

\cortext[cor1]{Corresponding author}

\address[LiPhy]{Université Grenoble Alpes, CNRS, LIPHY, 38000 Grenoble, France}
\begin{abstract}
The mechanism by which cells measure the dimension of the organ in which they are embedded, and slow down their growth when the final size is reached, is a long-standing problem of developmental biology. The role of mechanics in this feedback is considered important. Morphoelasticity is a standard continuum framework for modeling growing elastic tissues. However, in this theory, in the absence of additional variables, the feedback between growth and mechanical stress leads to either a collapse or unbounded growth of the tissue, but usually prohibits reaching a finite asymptotic size (‘size control’). In this article, we modify this classical setting to include an energetic cost associated with growth, leading to the physical effect of size control. The present model simultaneously provides a qualitatively correct residual stress profile and has a naturally emerging necrotic core, all of which have previously been experimentally established in multicellular spheroids. This is achieved through a local feedback mechanism derived from a thermodynamical framework. The model delivers testable predictions for experimental systems and could be a step towards the understanding of the role of mechanics in the multifaceted question of how growing organs attain their final size. 

\end{abstract}

\end{frontmatter}{}

\section{\label{sec:Introduction}Introduction}

In morphogenesis, living tissues change shape and size very rapidly. Morphogenetic events include spectacular shape changes such as self-inversion \cite{hohn2015dynamics}, looping as in the case of the heart tube \cite{ramasubramanian2008modeling}, branching such as in lungs, kidney and vascular networks \cite{erlich2019physical}. As these examples illustrate, morphogenesis involves complex interactions between growth, non-linear mechanics, shape and size, calling for mathematical approaches that can model such interplay. 

The determination of the appropriate form of evolution equations for growth and shape change has been the focus of much research (\cite{epstein2000thermomechanics,lubarda2002mechanics,dicarlo2002growth,ambrosi2007growth,ganghoffer2010mechanical}). With limited experimental data available, this research has employed thermodynamic arguments to motivate appropriate forms of the growth law that satisfy a dissipation inequality. The reasoning is typically to assume a single constituent theory in which every material point is in contact with a mass reservoir at an imposed chemical potential setting the tendency to grow. When the free energy depends only on the elastic deformation, by following a standard set of arguments and derivations, one arrives at a variant of the growth law
\begin{equation}
\dot{\mathbb{G}}\mathbb{G}^{-1}=\mathbb{K}(\mathbb{S}^{*}-\mathbb{S})\label{eq:generic-growth-law}
\end{equation}
where $\mathbb{G}$ is a tensor representing the geometrical rearrangement of the material due to growth, $\mathbb{S}$ is an Eshelby stress, $\mathbb{S}^{*}$ is a homeostatic stress representing the mass reservoir, and $\mathbb{K}$ is a constant symmetric positive-definite matrix of growth rates. In particular, the hypothesis behind mechanical homeostasis \cite{ambrosi2019growth,latorre2019mechanobiological,erlich2019homeostatic,goriely2017mathematics} is that in the state of homeostatic stress $\mathbb{S}=\mathbb{S}^{*}$, growth and shape change do not occur ($\dot{\mathbb{G}}=0$), as the cellular processes of birth, death and rearrangement balance each other out. Experimental evidence and physical understanding of homeostatic stress have been demonstrated in arteries, where residual stress is used to homogenize transmural stresses under physiological loading to minimize tissue abrasion during its lifetime \cite{chuong1986residual}. There is experimental evidence that some living systems, such as embryos \cite{beloussov2003morphomechanics} and fibroblast cells \cite{ezra2010changes}, maintain such a target, or homeostatic, stress.

The debate on the respective roles of mechanics and biochemistry in the study of how biological tissues control their size has undergone interesting developments in recent decades. How biological tissues control their size has been a decade-old mystery in mostly pure developmental biology research \cite{travis2013mysteries}. Many hypotheses have been experimentally tested and proven wrong: Size is not determined, for example, solely by a cell clock or cell counting \cite{vollmer2017growth}. As models based on reaction and diffusion of growth-promoting chemicals (morphogens) failed to explain numerous experimental observations \cite{day2000measuring}, mechanics has received increasing interest as a likely candidate for growth regulation in the developmental biology community \cite{shraiman2005mechanical,hufnagel2007mechanism,aegerter2007model,aegerter2012integrating}. For example, mechanics is fully accepted as a key ingredient in the growth of multicellular spheroids, a lab-made model system for the early stages of tissue expansion. Growth is affected by both the level of compression of the spheroid and the availability of oxygen \cite{stylianopoulos2012causes,ambrosi2004role,gao2016embryo,dolega2020mechanical}. An appealing aspect of the model \eqref{eq:generic-growth-law} it that it shifts the emphasis from biochemistry to mechanics, allowing a more nuanced understanding of the role of mechanics and residual stress in tissue growth and regulation, while still allowing a coarse-grained description of chemistry to enter through the external chemical potential $\mathbb{S}^*$ that can be refined as needed by, for instance, coupling growth to a diffusion process \cite{ambrosi2007growth}.

Certain details of the form of the growth law \eqref{eq:generic-growth-law} are not entirely agreed upon. The presence of Eshelby stress as a driving force of growth has been widely employed \cite{epstein2000thermomechanics,Ambrosi2005,ambrosi2007growth,ganghoffer2010mechanical,gao2016embryo}. The Eshelby stress was originally introduced to describe point forces due to elastic singularities, due to dislocations in crystal lattices \cite{eshelby1951force,eshelby1957determination}. It emerges naturally in the context of growth \cite{Ambrosi2005}. However, some authors make the assumption that the free energy of the incoming material matches the free energy of the pre-existing material, which leads to the presence of Mandel or Cauchy stress instead of Eshelby stress in the growth law \eqref{eq:generic-growth-law} (\cite{goriely2017mathematics,taber2008theoretical,taber1996theoretical}). The difference of the two approaches has been contrasted in \cite{buskohl2014influence}. In a similar vein, there is some disagreement on the form of the coefficient $\mathbb{K}$ in \eqref{eq:generic-growth-law}. A thermodynamical treatment  requires this coefficient to be a positive semi-definite matrix to ensure that the dissipation inequality is satisfied  \cite{Ambrosi2005}. However, certain authors who emphasize the role of unknown biochemical processes in thermodynamical treatments choose fourth-order coefficient tensors instead \cite{taber2009towards,goriely2017mathematics,erlich2019homeostatic}. This allows for cross-couplings in the growth dynamics that would be impossible in the classical treatment \cite{Ambrosi2005}, such as (in a system with spherical symmetry) the radial growth rate being coupled to hoop stress. 

A number of issues with \eqref{eq:generic-growth-law} have received relatively little attention: 
\begin{enumerate}
\item The first point concerns the homeostatic state itself. A conceptual problem raised by \eqref{eq:generic-growth-law} is that if at the tissue boundary the homeostatic stress does not match the boundary condition (which might be a prescribed hydrostatic pressure for instance), growth never stops at the boundary, making an equilibrium impossible. Several authors found a way around this problem by hypothesizing further evolution equations for $\mathbb{S}^{*}$ which adapt in a delayed response to the boundary conditions of the system (\cite{taber2008theoretical,taber2009towards}), or by postulating that the homeostatic stress is compatible with boundary conditions (\cite{erlich2018mechanical}). However, such choices are rather arbitrary in the context of biological tissue growth, where it is unclear why the reservoir of nutrients should be linked to the imposed mechanical boundary conditions.

\item The second point concerns a larger question in biology: Is the purely mechanical feedback mechanism \eqref{eq:generic-growth-law} sufficient to encode a final asymptotic size of the tissue? This ties into a larger debate in biology about how cells in an organ know what overall size the organ has, and how they ``decide'' when to stop dividing once the organ has reached the right size, and which role mechanics plays in such regulation (\cite{buchmann2014sizing,eder2017forces,vollmer2017growth,irvine2017mechanical,gou2020growth}). It is questionable whether the system \eqref{eq:generic-growth-law} will reach the same size or not depending on different initial conditions: \cite{ambrosi2015active,pettinati2016finite} hypothesized that it should not, but gave no proof or numerical example. While there have been some recent studies investigating the dynamics of \eqref{eq:generic-growth-law} with methods of dynamical systems theory (\cite{vandiver2009morpho,erlich2019homeostatic,latorre2019mechanobiological}), to our knowledge no investigation of the final size exists to date. 
\end{enumerate}

In this paper we will address both issues of \eqref{eq:generic-growth-law} by proposing a modification of the standard approach that overcomes the two issues mentioned. The idea is to penalize the growth process in the free energy of the system. While the classical theory neglects that growth has an energetic cost, even if the cell material building blocks are readily available in the extracellular fluid, it costs energy \cite{Alberts2002,phillips2009feeling} to get them through the cell membrane and assemble or disassemble them into the solid cell components that constitute the cell dry mass \cite{zhou2009universal}. An energetic cost is also involved in the ion pumping mechanism that is necessary to control cell volume and screen out the apparent osmotic imbalance between the cell inside and outside due to the presence of the macromolecules trapped in the cell \cite{cadart2019physics}, generally leading to a certain level of control of the cell mass density during growth \cite{grover2011measuring}. We show that using this concept, there is no longer a need for the homeostatic stress to match or adapt to the boundary conditions. Further, we show that this model creates the final size robustly, independent of variations in initial conditions, something which the classical model \eqref{eq:generic-growth-law} does not achieve, as we shall also demonstrate numerically. The final model qualitatively matches important known experimental observations about growing multicellular spheroids, namely that they reach an asymptotic size in the presence of external pressure \cite{helmlinger1997solid,alessandri2013cellular},  that the residual hoop stress near the periphery of the spheroid is tensile as consistent with cutting experiments \cite{stylianopoulos2012causes,colin2018experimental,guillaume2019characterization}, and that larger spheroids experience an inflow of material towards the core, known as a \emph{necrotic core} \cite{franko1979oxygen,delarue2013mechanical}. While mechano-chemical models captured some of these experimental observations \cite{ambrosi2017solid,xue2016biochemomechanical,walker2022towards}, to our knowledge, no existing model matches all these observations simultaneously. 

We divide the manuscript as follows. In Section \ref{sec:Kinematics-balance-laws-thermodynamics}, we introduce our notation and state the balance laws and the first and second principle of thermodynamics of a growing solid in finite elasticity. Our main idea, which distinguishes this work from classical works like \cite{rodriguez1994stress} and \cite{Ambrosi2005}, is that the free energy depends on the determinant of the growth tensor, $\left|\mathbb{G}\right|$, in addition to the traditional dependence on the elastic deformation gradient $\mathbb{A}$. In Section  \ref{sec:growth-law-with-energy-cost}, we propose a concrete form of this free energy to derive a growth law which offers a crucial modification compared to the classical law \eqref{eq:generic-growth-law}. The consequences of our modification are explored in Section \ref{sec:applications} based on two examples. Firstly, in subsection \ref{sec:1d-bar}, we consider a compressible uniaxially growing neo-Hookean bar. We find that our growth law enables the possibility of size control, and removes the need to prescribe the homeostatic pressure in the bar ad hoc, thus fixing both issues with the classical law discussed in the introduction. Secondly, in Section \ref{sec:compressible-spheroid} we consider a growing compressible neo-Hookean spheroid, showing how in addition to size control, residual stress can also be built up in this system through growth. A residual stress profile consistent with experiments can be controlled by the anisotropy of the homeostatic stress tensor $\mathbb{S}^*$. In Section \ref{sec:residual-stress-and-size-regulation}, we explore how the anisotropy influences the spheroid's ability to achieve size control. Large spheroids are subject to a flow of material towards the center, and we demonstrate how the necrotic core forms in our model in Section  \ref{sec:necrotic-core}. Finally, in section \ref{sec:Discussion}, we round up this work with a discussion. 

\begin{figure}[t]
\hfill{}\includegraphics{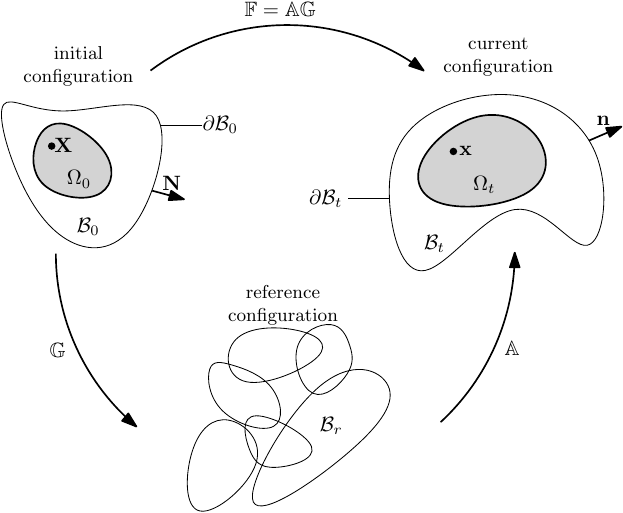}\hfill{}
\caption{\label{fig:relationships} Kinematic decomposition in morphoelasticity. The \emph{initial configuration} $\mathcal{B}_0$ describes the body in its initial state at time $t=0$ (i.e. before deformation and before growth) and is required to be stress-free. The growth tensor $\mathbb{G}$ describes growth without stress, leading to a stress-free  incompatible \emph{reference configuration} $\mathcal{B}_{r}$. The elastic deformation gradient $\mathbb{A}$ restores compatibility by introducing residual stress in the \emph{current configuration} $\mathcal{B}_{t}$. Material points in $\mathcal{B}_{0}$ are described by the vector $\mathbf{X}$ which is mapped to $\mathbf{x}$ in $\mathcal{B}_{t}$. Surface normals are denoted $\mathbf{N}$ on $\partial \mathcal{B}_0$ and $\mathbf{n}$ on $\partial \mathcal{B}_t$, respectively. Mass balance is stated over a subregion $\Omega_0 \subset \mathcal{B}_0$ in the initial configuration, which transforms to $\Omega_t \subset \mathcal{B}_t$ in the current configuration.}
\end{figure}

\section{\label{sec:Kinematics-balance-laws-thermodynamics}Kinematics, balance laws and thermodynamics}

\subsection{Kinematics}

From a mechanical perspective, a biological tissue of mammalian cells is typically constituted of cells interconnected directly by protein bonds or via some extracellular matrix. This system is permeated by an extracellular solvent which contains the nutrients and the building blocks necessary for its growth through biosynthesis and swelling of the cells followed by their division when reaching a critical added mass \cite{cadart2019physics}. We model the tissue as a single continuum which initial material points are parametrized by their position $\mathbf{X}\in\mathcal{B}_{0}$. As time evolves, the growth process is described by a two steps evolution of this initial state. The first step describes the growth of the \emph{initial configuration} $\mathcal{B}_{0}$ to a fictitious stress-free state, which we call the \emph{reference configuration} $\mathcal{B}_{r}$. The second step is the elastic mechanical response during which the tissue is deformed from the reference configuration to its actual deformed and stressed state, the \emph{current configuration}, $\mathbf{x}\in\mathcal{B}_{t}$, but no mass is added.

The  deformation map is $\boldsymbol{\varphi}:\,\,\mathcal{B}_{0}\rightarrow\mathcal{B}_{t}$, $\mathbf{X}\mapsto\mathbf{x}=\boldsymbol{\varphi}\left(\mathbf{X},t\right)$. The velocity of a material point $\mathbf{X}$ is $\mathbf{v}=\dot{\boldsymbol{\varphi}}\left(\mathbf{X},t\right)$ where the overdot denotes the material time derivative at fixed material coordinate $\mathbf{X}$. Denoting $\mathbb{G}$ the tensor describing the change of material configuration due to growth and $\mathbb{A}$ the one due to the material elastic response that makes it compatible, the deformation gradient  $\mathbb{F}=\nabla_{\mathbf{X}}\boldsymbol{\varphi}\left(\mathbf{X},t\right)$ reads \cite{rodriguez1994stress}: 
\begin{equation}
\mathbb{F}=\frac{\partial\mathbf{x}}{\partial\mathbf{X}}=\mathbb{A}\mathbb{G}.\label{transf_grad}
\end{equation}
The morphoelastic decomposition is illustrated in Fig. \ref{fig:relationships}. Physically, $\mathbb{G}$  leads in general to an incompatible configuration of the body ($\mathcal{B}_r$ is said not to fit into Euclidean space and therefore sketched in Fig. \ref{fig:relationships} with holes and overlaps) \cite{truskinovsky2019nonlinear,goriely2017mathematics}. For this reason, the current configuration $\mathcal{B}_{t}$ is no longer stress free, even when it is unloaded. Stress that remains even in the absence of loads is called \textit{residual stress}.

Further, we denote the right Cauchy-Green strain tensor as $\mathbb{C}=\mathbb{A}^{\mathsf{T}}\mathbb{A}$ and the left Cauchy-Green tensor as $\mathbb{B}=\mathbb{A}\mathbb{A}^{\mathsf{T}}$. 

\subsection{\label{subsec:Mass-balance-generic}Mass balance}

 In our continuum modeling approach, the tissue has a density (mass per volume) $\rho\left(\mathbf{X},t\right)$ in $\mathcal{B}_{t}$. The biosynthesis and subsequent swelling and division of the cells   contributes a volumetric growth rate function $\rho\left(\mathbf{X},t\right)\Gamma\left(\mathbf{X},t\right)$ in $\mathcal{B}_{t}$. Physical laws are most naturally stated in the current configuration. The mass balance takes the form
\begin{equation}
\dot{\overline{\int_{\Omega_{t}}\rho\,\mathrm{d}\mathbf{x}}}=\int_{\Omega_{t}}\rho\Gamma\,\mathrm{d}\mathbf{x}.\label{eq:mass-balance-generic}
\end{equation}
 The infinitesimal volume element of a material point $\mathbf{X}$ in initial and current configuration are, respectively, $\mathrm{d}\mathbf{X}$ and $\mathrm{d}\mathbf{x}$. They transform via the Jacobian $\left|\mathbb{F}\right|$, which represents the local change of volume, that is $\mathrm{d}\mathbf{x}=$$\left|\mathbb{F}\right|\mathrm{d}\mathbf{X}$. 

Mass conservation between the reference configuration $\mathcal{B}_{r}$ and current configuration $\mathcal{B}_{t}$ leads to the relationship 
\begin{equation}
\rho_{r}=\rho\,\left|\mathbb{A}\right|\,.\label{eq:reference-density-definition}
\end{equation}
We further assume that the reference density is a constant in time, $\dot{\rho}_{r}=0$. Taking into account this assumption and transforming the mass balance \eqref{eq:mass-balance-generic} into the initial configuration, we get $\dot{\overline{\int_{\Omega_{t}}\rho\,\mathrm{d}\mathbf{x}}}=\int_{\Omega_{0}}\left|\mathbb{F}\right|\rho\Gamma\,\mathrm{d}\mathbf{X}\,.$ Assuming that the integrands are all continuous and using the fact that $\Omega_{t}$ is arbitrary, we use the \emph{Maxwell transport }and \emph{localization procedure} (\cite{gurtin2010mechanics,goriely2017mathematics}) to obtain the local version of the mass balance equation: $\dot{\overline{\left|\mathbb{F}\right|\rho}}=\rho_{r}\dot{\left|\mathbb{G}\right|}=\rho|\mathbb{F}|\text{tr}(\dot{\mathbb{G}}\mathbb{G}^{-1}).$ Here we applied Jacobi's identity $\dot{\left|\mathbb{G}\right|}=\left|\mathbb{G}\right|\text{tr}(\mathbb{G}^{-1}\dot{\mathbb{G}})$ (\cite{gurtin2010mechanics}). That allows us to identify the mass source term as
\begin{equation}
\Gamma=\text{tr}(\dot{\mathbb{G}}\mathbb{G}^{-1}).
\end{equation}
The trace, and double contraction operator, are defined in \ref{sec:Explicit-form-of-dfdA-dfdG}.

\subsection{Momentum balance}

In the absence of external body force and inertia, we write momentum balance as $\nabla_{\mathbf{x}}\cdot\mathbb{T}=0$ where $\mathbb{T}$ is the Cauchy stress tensor. The angular momentum balance imposes symmetry of the Cauchy stress tensor $\mathbb{T}^{\mathsf{T}}=\mathbb{T}$. 

\subsection{First and second principle}

Combining the first and second principles of thermodynamics, at a fixed temperature, the dissipation $\Theta$ (i.e. the entropy production rate divided by the temperature) takes a simple form $\Theta=\dot{\mathcal{E}}+\dot{\mathcal{W}}-\dot{\mathcal{F}}\geq0$, where $\mathcal{W}$ is the external mechanical work performed on the tissue through its boundary, $\mathcal{F}=\mathcal{U}-\mathcal{T}\mathcal{S}$ is the Helmholtz free energy of the tissue ($\mathcal{U}$ is the internal energy, $\mathcal{S}$ the entropy, and $\mathcal{T}$ the temperature) and $\mathcal{E}$ is the free energy exchanged with the external environment. 

We assume that mass is delivered within the tissue bulk such that:
\begin{equation}
\dot{\mathcal{E}}=\int_{\mathcal{B}_{t}}\rho\mathbb{S}^{*}:\dot{\mathbb{G}}\mathbb{G}^{-1}\mathrm{d}\mathbf{x},\label{eq:external-free-energy}
\end{equation}
where $\mathbb{S}^{*}$ is an imposed tensorial chemical potential \cite{truskinovskiy1983chemical} setting the building blocks availability in the local environment along the growth principal directions, sometimes referred to as the homeostatic stress tensor \cite{Ambrosi2005,taber2009towards,buskohl2014influence,erlich2019homeostatic}. 

 The mechanical work is done through traction forces $\mathbf{t}=\mathbb{T}\mathbf{n}$ acting at the boundary $\partial \mathcal{B}_t$ with surface normal $\mathbf{n}$ (see Fig. \ref{fig:relationships} for an illustration). Using the momentum balance $\nabla_{\mathbf{x}}\cdot\mathbb{T}=0$, the rate of work is $\dot{\mathcal{W}}=\int_{\mathcal{B}_{t}}\mathbb{T}:\nabla_{\mathbf{x}}\mathbf{v}\,\mathrm{d}\mathbf{x}\,.$ Finally, we write the free energy as $\mathcal{F}=\int_{\mathcal{B}_{t}}\rho f\,\mathrm{d}\mathbf{x}$ in terms of the free energy per current unit mass $f$, of which the form specific will be detailed in the next few sections. Transformed into the initial configuration, the dissipation inequality reads 
\begin{equation}
\Theta=\underbrace{\int_{\mathcal{B}_{0}}\left|\mathbb{F}\right|\rho\mathbb{S}^{*}:\dot{\mathbb{G}}\mathbb{G}^{-1}\,\mathrm{d}\mathbf{X}}_{\dot{\mathcal{E}}}+\underbrace{\int_{\mathcal{B}_{0}}\left|\mathbb{F}\right|\mathbb{T}:\nabla_{\mathbf{x}}\mathbf{v}\,\mathrm{d}\mathbf{X}}_{\dot{\mathcal{W}}}-\underbrace{\int_{\mathcal{B}_{0}}\dot{\overline{\left|\mathbb{F}\right|\rho f}}\,\mathrm{d}\mathbf{X}}_{\dot{\mathcal{F}}}\geq0\,.\label{eq:dissipation-inequality-generic}
\end{equation}

We assume that the free energy density $f$ depends on the elastic deformation $\mathbb{A}$ and the determinant of the growth tensor $\left|\mathbb{G}\right|$, that is $f=f\left(\mathbb{A},\left|\mathbb{G}\right|\right)\,.$ 
While the dependence in $\mathbb{A}$ corresponds to the classical morphoelastic  theory \cite{rodriguez1994stress,goriely2017mathematics}, the $\left|\mathbb{G}\right|$ dependence reflects the fact that, during the cell cycle, the swelling  of the cell volume by the uptake of extracellular fluid costs some metabolic energy (e.g. biosynthesis of proteins, ion pumps activity). Over a longer timescale, cell division also requires some energy, for instance for the  active process of cytokinesis to take place. With this assumption, the dissipation inequality \eqref{eq:dissipation-inequality-generic} can be rewritten as 

\begin{equation}
\Theta=\int_{\mathcal{B}_0}\left(\left|\mathbb{F}\right|\mathbb{T}\mathbb{A}^{-\mathsf{T}}-\left|\mathbb{F}\right|\rho\frac{\partial f}{\partial\mathbb{A}}\right):\dot{\mathbb{A}}+\left|\mathbb{F}\right|\rho\underbrace{\left(\mathbb{S}^{*}+\frac{1}{\rho}\mathbb{A}^{\mathsf{T}}\mathbb{T}\mathbb{A}^{-\mathsf{T}}-f\,\mathds{21}-\frac{\partial f}{\partial\left|\mathbb{G}\right|}\left|\mathbb{G}\right|\mathds{21}\right)}_{\boldsymbol{\mathcal{G}}}:\dot{\mathbb{G}}\mathbb{G}^{-1}\,\mathrm{d}\mathbf{X}\geq0\label{eq:Dissipation-over-Jg}
\end{equation}
Since the rates $\dot{\mathbb{A}}$ and $\dot{\mathbb{G}}$ can vary independently, this inequality must hold for each term independently. The assumption that stress is non-dissipative (elastic mechanical response) thus enforces:
\begin{equation}
    \mathbb{T} =\rho\frac{\partial f}{\partial\mathbb{A}}\mathbb{A}^{\mathsf{T}}\,.\label{eq:Cauchy-stress-psi}
\end{equation}
We define the elastic energy per unit volume of the reference configuration $W$ as
\begin{equation}
f\left(\mathbb{A},\left|\mathbb{G}\right|\right)=\frac{1}{\rho_{r}}W\left(\mathbb{A},\left|\mathbb{G}\right|\right).\label{eq:free-energy-in-reference-config}
\end{equation}
The Cauchy stress \eqref{eq:Cauchy-stress-psi} then takes the form
\begin{equation}
\mathbb{T}=\frac{1}{\left|\mathbb{A}\right|}\frac{\partial W}{\partial\mathbb{A}}\mathbb{A}^{\mathsf{T}}\,.\label{eq:Cauchy-stress-W}
\end{equation}
We can see that the dissipation inequality simplifies by inserting \eqref{eq:Cauchy-stress-psi} into \eqref{eq:Dissipation-over-Jg}. This leaves 
$\Theta=\int_{\mathcal{B}_0}\left|\mathbb{G}\right|\rho_{r}\mathcal{G}:\dot{\mathbb{G}}\mathbb{G}^{-1}\mathrm{d}\mathbf{X}\geq0$. To ensure that the dissipation inequality is satisfied, we assume the close-to-equilibrium relation $\mathbb{K}\mathcal{G}=\dot{\mathbb{G}}\mathbb{G}^{-1}$ where $\mathbb{K}$ is a constant symmetric positive-definite matrix. We thus obtain the growth law
\begin{equation}
\dot{\mathbb{G}}\mathbb{G}^{-1}=\mathbb{K}\left(\mathbb{S}^{*}-f\,\mathds{21}+\frac{1}{\rho}\mathbb{A}^{\mathsf{T}}\mathbb{T}\mathbb{A}^{-\mathsf{T}}-\frac{\partial f}{\partial\left|\mathbb{G}\right|}\left|\mathbb{G}\right|\mathds{21}\right).
\label{eq:growth-law-TD-generic}
\end{equation}
Further, we identify the ``Eshelby-like'' stress tensor \cite{ambrosi2007growth,Ambrosi2005} $\mathbb{S}$: 
\begin{equation}
\mathbb{S}:=f\,\mathds{21}-\mathbb{A}^{\mathsf{T}}\frac{\partial f}{\partial\mathbb{A}}=f\,\mathds{21}-\frac{1}{\rho}\mathbb{A}^{\mathsf{T}}\mathbb{T}\mathbb{A}^{-\mathsf{T}}=\frac{1}{\rho_{r}}\left(W\mathds{21}-\left|\mathbb{A}\right|\mathbb{A}^{\mathsf{T}}\mathbb{T}\mathbb{A}^{-\mathsf{T}}\right)\,.\label{eq:Eshelby-stress-definition}
\end{equation}
 Next, we non-dimensionalize the system, using the following scaling and notation in which the hatted variables are dimensionless: 
\begin{equation}
\widehat{t}=\frac{t}{\tau},\quad\hat{\dot{\overline{\left(\ldots\right)}}}=\frac{\partial\left(\ldots\right)}{\partial\widehat{t}},\quad\left\{ \widehat{\mathbb{S}},\widehat{\mathbb{S}}^{*},\widehat{f}\right\} =\frac{\rho_{r}}{G}\left\{ \mathbb{S},\mathbb{S}^{*},f\right\} ,\quad\left\{ \widehat{\mathbb{T}},\widehat{W}\right\} =\frac{1}{G}\left\{ \mathbb{T},W\right\} ,\quad\widehat{\mathbb{K}}=\frac{\tau G}{\rho_{r}}\mathbb{K}\,.
\label{eq:scaling}
\end{equation}
where $\tau$ is the characteristic time scale of growth. This leads to the dimensionless growth law 
\begin{equation}
\hat{\dot{\mathbb{G}}}\mathbb{G}^{-1}=\widehat{\mathbb{K}}\left(\widehat{\mathbb{S}}^{*}-\widehat{\mathbb{S}}-\frac{\partial\widehat{f}}{\partial\left|\mathbb{G}\right|}\left|\mathbb{G}\right|\mathds{21}\right)
\label{eq:growth-law-non-dim}
\end{equation}
where the non-dimensional Eshelby stress is $\widehat{\mathbb{S}}=\widehat{W}\mathds{21}-\left|\mathbb{A}\right|\mathbb{A}^{\mathsf{T}}\widehat{\mathbb{T}}\mathbb{A}^{-\mathsf{T}}$. 

We now make another simplifying assumption regarding $\widehat{\mathbb{K}}$ in  \eqref{eq:growth-law-non-dim} assuming that the rate of growth is isotropic such that $\widehat{\mathbb{K}}=\widehat{k}\mathds{21}$.
 Since $\widehat{k}$ is used to non-dimensionalized time, without loss of generality,  we will set $\widehat{\mathbb{K}}=\mathds{21}$ for the rest of this manuscript.

\section{\label{sec:growth-law-with-energy-cost}The energy cost due to growth}

We now consider the following expression of the  energy density $W$,
\begin{equation}
W\left(\mathbb{A},\left|\mathbb{G}\right|\right)=W_{\text{el}}\left(\mathbb{A}\right)+W_{\text{g}}\left(\left|\mathbb{G}\right|\right)=\frac{G}{2}\left(I_{1}-3-2\log\left|\mathbb{A}\right|\right)+\frac{\kappa}{2}\left(\left|\mathbb{A}\right|-1\right)^{2}+\frac{\chi}{2}\left(\frac{\left|\mathbb{G}\right|-1}{\left|\mathbb{G}\right|+1}\right)^{2}.
\label{eq:quadratic-form-free-energy}
\end{equation}
The first part of the energy, $W_\text{el}$,  is that of a compressible neo-Hookean material \cite{pence2015compressible}, where $G$ is the shear modulus and $\kappa$ the bulk modulus. The last term, $W_\text{g}$, weighted by the growth penalty $\chi$,  is an energetic penalty due to growth, which is not present in most current treatments of growth \cite{epstein2000thermomechanics,lubarda2002mechanics,dicarlo2002growth,Ambrosi2005,ganghoffer2010mechanical,buskohl2014influence}, \chg{although a dependence of this type has been suggested in \cite{cyron2017mechanobiological}}. We non-dimensionalize $\kappa$ and $\chi$ according to 
\begin{equation}
\left\{ \widehat{\kappa},\widehat{\chi}\right\} =\frac{1}{G}\left\{ \kappa,\chi\right\} \,.
\label{eq:scaling-part-2}
\end{equation}
Taking also into account \eqref{eq:scaling}, the energy density \eqref{eq:quadratic-form-free-energy} in non-dimensional form becomes
\begin{equation}
\widehat{W}\left(\mathbb{A},\left|\mathbb{G}\right|\right)=\widehat{W}_{\text{el}}\left(\mathbb{A}\right)+\widehat{W}_{\text{g}}\left(\left|\mathbb{G}\right|\right)=\frac{1}{2}\left(I_{1}-3-2\log\left|\mathbb{A}\right|\right)+\frac{\widehat{\kappa}}{2}\left(\left|\mathbb{A}\right|-1\right)^{2}+\frac{\widehat{\chi}}{2}\left(\frac{\left|\mathbb{G}\right|-1}{\left|\mathbb{G}\right|+1}\right)^{2}.
\label{eq:quadratic-form-free-energy-nondim}
\end{equation}

We shall motivate the new $\widehat{W}_\text{g}$ term in two steps. In Section \ref{sec:no-mechanical-feedback}, we focus in isolation on the $\widehat{W}_\text{g}$ term and demonstrate in a geometrically simplified setting how its presence in the free energy density leads to a growth dynamics with size control. In Section \ref{sec:full-growth-law}, we construct the fully general growth law pertaining to the free energy density \eqref{eq:quadratic-form-free-energy}, a growth law in which both the elastic term and the growth penalty term are present.  

\begin{figure}[t]
\hfill{}\includegraphics[width=0.55\textwidth]{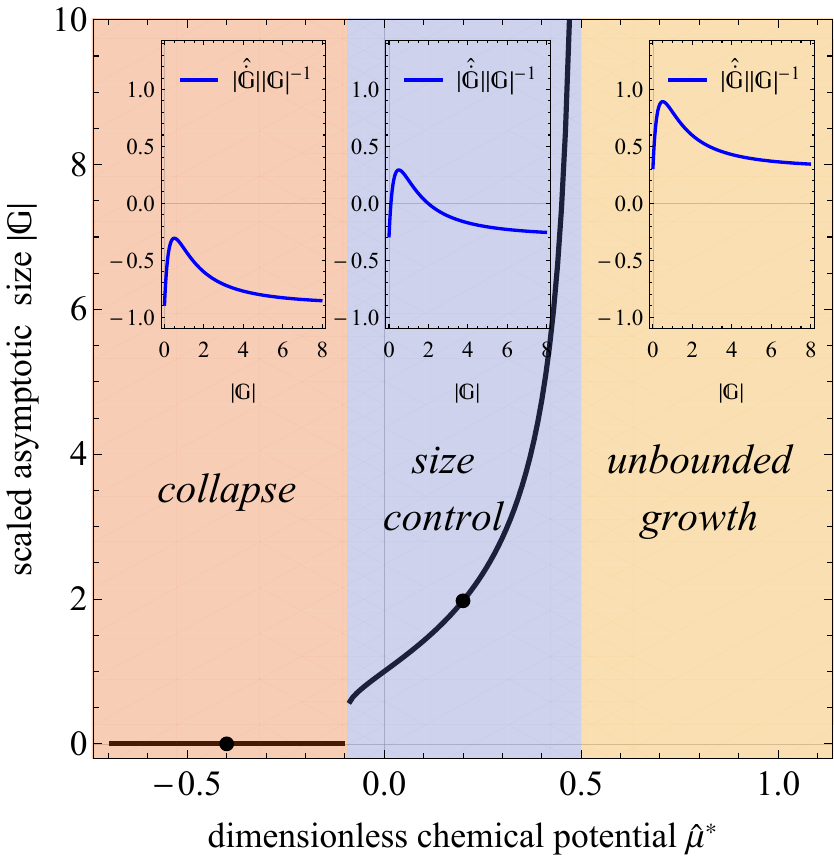}\hfill{}
\caption{Final size in the absence of mechanical feedback. Depending on the value of the scaled chemical potential $\widehat{\mu}^*$, the dynamics is in one of three regions: collapse, size control, and unbounded growth. The insets show the right hand side of the dynamical system \eqref{eq:growth-dynamics-no-MF}, for a representative point in the respective region (the points are marked as black circles). If the nutrient supply, modeled with the chemical potential $\widehat{\mu}^*$, is low enough, the cell suspension dies and reaches zero final volume. If $\widehat{\mu}^*$ is sufficiently high, on the other hand, the suspension grows without any control or limit. Between those regimes, a fixed final size $|\mathbb{G}|$ for $t\rightarrow \infty$ is reached, independently of the initial conditions of  \eqref{eq:growth-dynamics-no-MF}. This represents controlled growth of the cells, which we term \emph{size control}. \chg{Parameter value: $\widehat{\chi}=1$.} }  
\label{fig:no-mechanical-feedback}
\end{figure}

\subsection{\label{sec:no-mechanical-feedback}No mechanical feedback}

To motivate the last term of \eqref{eq:quadratic-form-free-energy}, \chg{$W_\text{g}$, }we first consider the case of a cell colony where mechanics is not playing any role (for instance when cells are isolated in suspension or on a petri dish). In this case, the colony still grows according to the nutrients availability (i.e. the carrying capacity of the environment) and there are usually three regimes. If the nutrients supply is large enough, growth will be exponential while we expect some growth arrest in an environment with moderate supplies. \chg{In the context of population dynamics, the exponential regime is referred to as Malthusian growth \cite{goriely2017mathematics}.  If the supply is too low, the colony will die (extinction in the context of population dynamics). Finally, if there is moderate nutrient supply, the population of cells will saturate in a logistic profile, referred to as Verhulst growth or Gompertzian growth \cite{roose2007mathematical}.  These models are frequently applied to cancer modeling \cite{jarrett2018mathematical}. Various classical models within the sigmoidal function family, such as the von Bertalanffy, Gompertzian, and logistic models, have been used to describe the growth of tumors \cite{vaidya1982evaluation,freyer1988role}. The concept of a "carrying capacity", borrowed from population dynamics and ecology, represents the maximum number of cells for a growing cell population in an environment with given supplies. \cite{gatenby2003evolutionary}.} 

This situation is mimicked when considering the growth law \eqref{eq:growth-law-non-dim} in the absence of mechanics (i.e., $f$ is independent of $\mathbb{A}$), which is what we would get if we set $G=0$ and $\kappa=0$ in \eqref{eq:quadratic-form-free-energy}. We denote this energy (in non-dimensional form) as  $\widehat{W}=\widehat{W}_\text{g}\left(\left|\mathbb{G}\right|\right)$, and its derivative $\mathrm{d}\widehat{W}_\text{g}/\mathrm{d}\left|\mathbb{G}\right|$ by $\widehat{W}_\text{g}'\left(\left|\mathbb{G}\right|\right)$. We next assume that the homeostatic stress is spherical, $\widehat{\mathbb{S}}^{*}=\widehat{\mu}^{*}\mathds{21}$, where we $\widehat{\mu}^{*}$ is an external chemical potential. Furthermore, we assume that $\mathbb{G}$ is spherical, that is $\mathbb{G}=\left|\mathbb{G}\right|^\frac{1}{3}\mathds{1}$, in which case $\mathds{1}:\hat{\dot{\mathbb{G}}}\mathbb{G}^{-1}=|\hat{\dot{\mathbb{G}}}||\mathbb{G}|^{-1}$. Then the non-dimensional growth law \eqref{eq:growth-law-non-dim} simplifies to  
\begin{equation}
|\hat{\dot{\mathbb{G}}}||\mathbb{G}|^{-1}=\widehat{\mu}^{*}-\widehat{W}_{g}-\left|\mathbb{G}\right|\widehat{W}_{g}'\left(\left|\mathbb{G}\right|\right)=\widehat{\mu}^{*}-\frac{\left(\left|\mathbb{G}\right|-1\right)\left(\left|\mathbb{G}\right|(\left|\mathbb{G}\right|+4)-1\right)\widehat{\chi}}{2\left(\left|\mathbb{G}\right|+1\right)^{3}}\,.
\label{eq:growth-dynamics-no-MF}
\end{equation}

In Fig. \ref{fig:no-mechanical-feedback}, we show how the availability of nutrients, represented by the chemical potential $\widehat{\mu}^*$, affects the asymptotic size of the system described by the dynamical equation \eqref{eq:growth-dynamics-no-MF}. When $\widehat{\mu}^*$ is lower than a threshold value, the cell suspension collapses to zero size (red region). The right hand side of the dynamical system \eqref{eq:growth-dynamics-no-MF}, plotted as an inset, has no zeros, and the dynamics has no equilibrium points. As $\widehat{\mu}^*$ is increased, we reach a finite size that increases with the value of $\widehat{\mu}^*$ (blue region). The right hand side of \eqref{eq:growth-dynamics-no-MF} has two zeros (inset), one of which is a stable equilibrium. We term this region "size control". Finally, once  $\widehat{\mu}^*$ becomes large enough to leave the size control region, we enter a regime of unbounded growth (yellow region). The right hand side of \eqref{eq:growth-dynamics-no-MF} once again has no zeros and is always positive, making the growth unbounded. 

\chg{The $W_\text{g}$ term implies the existence of a memory of a reference configuration in the cells of the tissue. Indeed, telomere length is known to keep track of the number of divisions that a given cell has undergone \cite{hayflick1965limited,shay2000hayflick}. Furthermore, there is a clear size memory at the cell scale as daughter cells always reach twice the volume of their mother cell before they divide  \cite{cadart2019physics}.}


\subsection{\label{sec:full-growth-law}Full growth law}

\chg{Next, our goal is to investigate how the classical population dynamics models couple with morphoelasticity.} To this end, we must evaluate the expression \eqref{eq:growth-law-non-dim} using the free energy \eqref{eq:quadratic-form-free-energy-nondim}. Since the Cauchy stress $\mathbb{T}$ appears in the growth law, it needs to be evaluated according to \eqref{eq:Cauchy-stress-W}, once again using the free energy \eqref{eq:quadratic-form-free-energy}.  For this, the expressions $\partial f/\partial\mathbb{A}$ and $\partial f/\partial\left|\mathbb{G}\right|$ are required. Their explicit computation is detailed in  \ref{sec:Explicit-form-of-dfdA-dfdG}, expressions \eqref{eq:dfdA} and \eqref{eq:dfdG}. Taking those results into account, the Cauchy stress is 
\begin{equation}
\mathbb{T}=\frac{G}{\left|\mathbb{A}\right|}\left(\mathbb{B}-\mathds{21}\right)+\kappa\left(\left|\mathbb{A}\right|-1\right)\mathds{21}\,\label{eq:Cauchy-stress-compressible}
\end{equation}
and the non-dimensional growth law \eqref{eq:growth-law-non-dim} reads
\begin{equation}
\hat{\dot{\mathbb{G}}}\mathbb{G}^{-1}=\widehat{\mathbb{S}}^{*}-\widehat{\mathbb{S}}-2\widehat{\chi}\left|\mathbb{G}\right|\left(\frac{\left|\mathbb{G}\right|-1}{\left(\left|\mathbb{G}\right|+1\right)^{3}}\right)\mathds{21}\,.
\label{eq:growth-law-compressible-nH}
\end{equation}

\subsection{\label{sec:analogy-strain-gradient-plasticity}\chg{Analogy between growth energy $W_\text{g}$ and defect energy in crystal plasticity}}

\chg{While most growth theories 
\cite{dicarlo2002growth,epstein2000thermomechanics,ambrosi2008stress,gao2016embryo,fraldi2018cells,ambrosi2007growth,ciarletta2012mass} do not take into account a free-energy contribution due to growth ($W_\text{g}$ in \eqref{eq:quadratic-form-free-energy}), in plasticity, it is common to include an anelastic contribution into the free energy. In crystal plasticity, the deformation gradient $\mathbb{F}=\mathbb{A}\mathbb{P}$ is decomposed into an elastic tensor $\mathbb{A}$ and a plastic tensor $\mathbb{P}$ of which $\mathbb{G}$ is the kinematic equivalent in growth theories.  The free energy then takes the form $W=W_{\text{el}}\left(\mathbb{A}\right)+W_{\text{d}}(\gamma_p)$ where $W_{\text{el}}$ is the retrievable energy connected to atomic bond stretching in a crystal lattice, and $W_{\text{d}}$ is a defect energy due to imperfections in the lattice such as  dislocations. 
 $W_{\text{d}}$ is often called the \textquoteleft \textquoteleft stored energy of cold work\textquoteright \textquoteright 
 \cite{gurtin2009thermodynamics,hurtado2013finite}. The defect energy is a function of $\gamma_p$, which is a macroscopic measure of the dislocations stored in the microscopic structure. This quantity is closely linked to the plastic deformation gradient: $\gamma_p$ is  the accumulated plastic strain defined in terms of the differential equation  $\dot{\gamma}_{p}=|\mathbb{D}_{p}|$ where the plastic stretching $\mathbb{D}_{p}$ is the symmetrization of the plastic distortion-rate tensor, that is $\mathbb{D}_{p}=\text{sym}(\dot{\mathbb{P}}\mathbb{P}^{-1})$ \cite{gurtin2009thermodynamics,gurtin2010mechanics,hurtado2013finite}. A extension of this view, strain gradient plasticity, further assumes that $W_{\text{d}}$ also depends on gradients of the accumulated plastic strain, which quantifies the inhomogeneity of the microscopic structure due to stored dislocations \cite{gurtin2002gradient,gurtin2005theory,leoni2010gradient,de2012gamma,ponsiglione2007elastic,cermelli2001characterization}. }

In both crystal plasticity and growth theories, incompatibility plays a very important role. While in plasticity the geometric incompatibility usually arises due to defects in the crystal lattice, the source of incompatibility in growth is as diverse as biology itself: Incompatibility may be due to isotropic non-uniform growth (for instance, in tumor spheroids \cite{ambrosi2017solid}), anisotropic uniform  growth (for instance in the \textit{Drosophila} wing disc \cite{harmansa2022growth}), or differential growth between adjacent layers  (for instance in  heart tube formation \cite{hosseini2017new} or chick eye development \cite{oltean2016tissue}).

\begin{figure}[t]
	\hfill{}\includegraphics[width=1\textwidth]{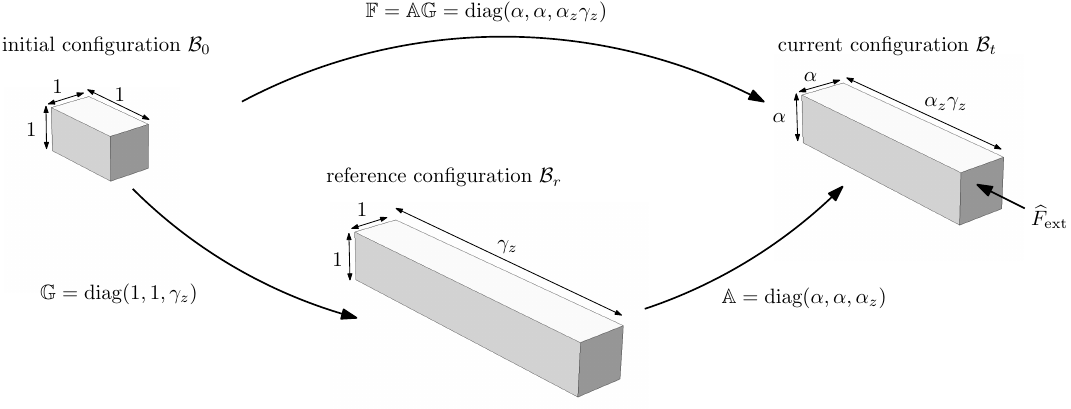}\hfill{}
	\caption{Illustration of a single bar growing uniaxially. In the initial configuration $\mathcal{B}_{0}$, before any growth or elasticity takes place, the system is stress-free. For the purpose of illustration, we set all reference lengths to 1. The growth stretches $\gamma_z$ describes the change of length purely due to growth in $z$-direction, where $\gamma_z$ in the reference configuration $\mathcal{B}_{r}$ can be thought of as rest-lengths of a (non-linear) spring. The elastic deformation with respect to these rest lengths is captured via the elastic stretches $\alpha_z$, where  $\lambda_z=\alpha_z \gamma_z$ in the current configuration $\mathcal{B}_{t}$ represent the observed (scaled) length of the bar, which is subject to an external force $\widehat{F}_\text{ext}$ acting over the (scaled) surface $\alpha^2$. }
	\label{fig:two-bars-in-parallel} 
\end{figure}

\section{\label{sec:applications}Application to a 1D and 3D scenario}

In this section, we explore the consequences of the modification to the free energy density \eqref{eq:quadratic-form-free-energy-nondim} and the growth law \eqref{eq:growth-law-compressible-nH} that follows from it. We start in subsection \ref{sec:1d-bar}, with a uniaxially growing bar, which is restricted to homogeneous deformations. In this system, deformations and stresses are spatially uniform, making the mathematical analysis transparent. Next, in Section \ref{sec:compressible-spheroid}, we study a growing spheroid. This system permits anisotropy in growth and stress, which leads to the buildup of residual stress. 

\subsection{\label{sec:1d-bar} 1D: Uniaxial growth of a single bar}

\begin{figure}[t]
\hfill{}\includegraphics[width=0.75\textwidth]{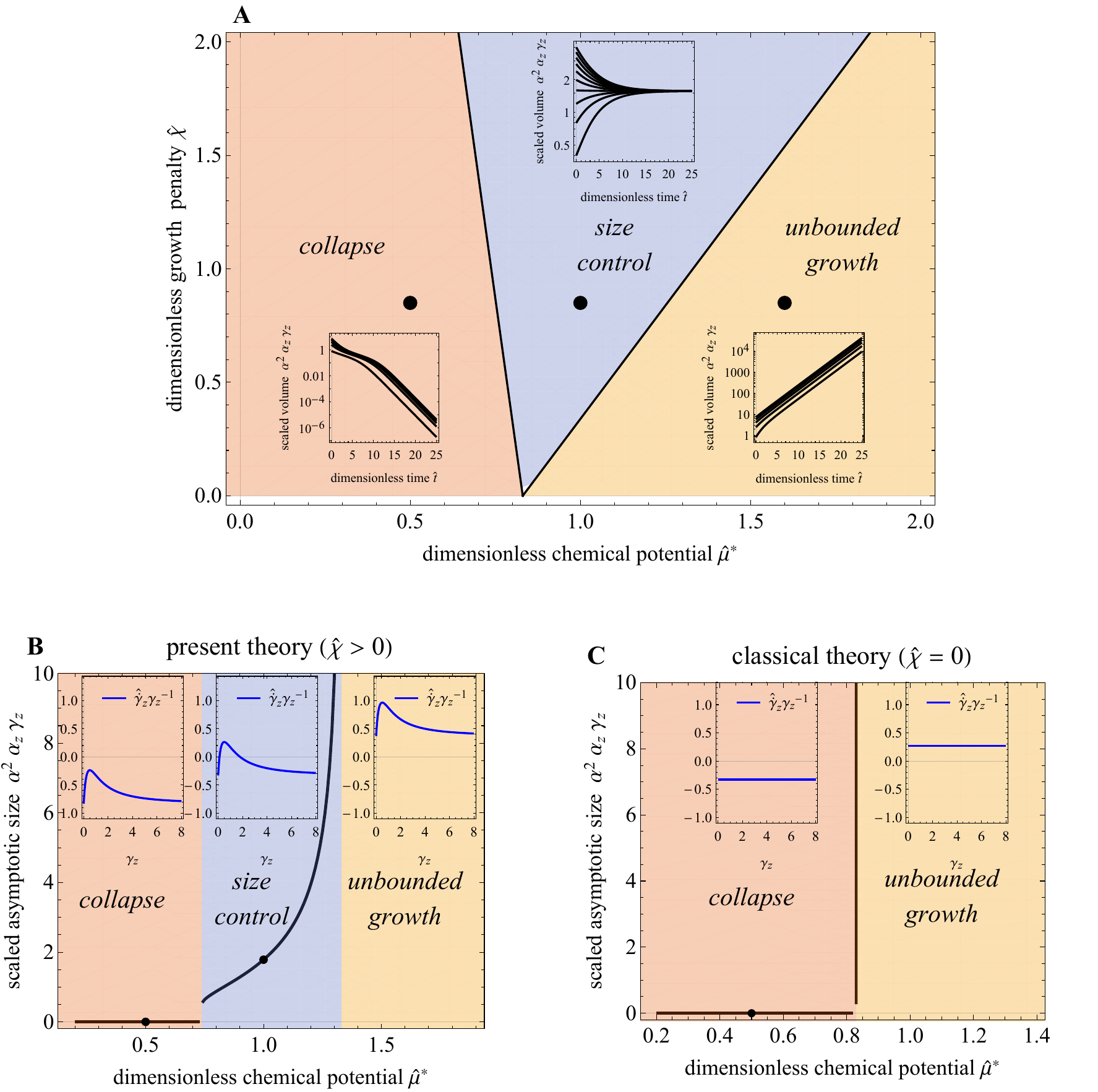}\hfill{}

\caption{Dynamics and final size of a single bar growing uniaxially. \textbf{A} Region diagram showing the asymptotic final size as a function of the non-dimensional parameters $\widehat{\mu}^*$ and $\widehat{\chi}$. The present theory "opens up" a blue region in which a finite non-zero final size is reached, independently of initial conditions (see trajectories inset, where the $\widehat{\mu}^*$ and $\widehat{\chi}$ parameters are marked by the black dot). This region is placed between the red "collapse" region, in which the system reaches a zero size, and the yellow "exponential growth" regime, in which the system keeps growing exponentially without bounds. The classical theory can be seen in the cusp of the "V" shaped region, in which case there is only collapse and unbounded growth, but no size control.   \textbf{B} The final asymptotic size, as well as the right-hand side of the growth law  \eqref{eq:bar-growth-law} (as insets). The final size is finite and non-zero in the size control region. \textbf{C} In the classical theory  ($\widehat{\chi}=0$)  the size control region disappears.  Parameters are $\widehat{\kappa}=1$, $\widehat{\mu}^{*}=1$, $\widehat{F}_\text{ext}=-1$. }
\label{fig:1D-model}
\end{figure}

We illustrate the presence and absence of size control based on the simplest possible geometry, a uniaxially growing bar, in which all deformations are spatially homogeneous. We work on a Cartesian coordinate basis $\left\{ \mathbf{e}_{x},\mathbf{e}_{y},\mathbf{e}_{z}\right\} $ in both initial and current configurations, with the direction of growth being $\mathbf{e}_{z}$. We shall write tensor components in this basis from now on. Assuming symmetric deformations in $\mathbf{e}_{x}$ and $\mathbf{e}_{y}$ directions, as well as uniaxial growth, the kinematic tensors are
\begin{equation}
\mathbb{F}=\text{diag}\left(\lambda,\lambda,\lambda_{z}\right),\qquad\mathbb{A}=\text{diag}\left(\alpha,\alpha,\alpha_{z}\right),\qquad\mathbb{G}=\text{diag}\left(1,1,\gamma_z\right)\,,
\end{equation}
see Fig. \ref{fig:two-bars-in-parallel}. The Cauchy stress is then given by \eqref{eq:Cauchy-stress-compressible}. Assuming no traction on the long faces of the bar, as well as a non-dimensional external force $\widehat{F}_{\text{ext}}$ distributed over the scaled area $\alpha^{2}$ of the remaining faces, the force boundary conditions are 
\begin{equation}
\widehat{T}_x=\widehat{T}_y=\frac{\left(\alpha^{2}-1\right)}{\alpha^{2}\alpha_{z}}+\widehat{\kappa}\left(\alpha^{2}\alpha_{z}-1\right)=0,\qquad \widehat{T}_z=\frac{\left(\alpha_{z}^{2}-1\right)}{\alpha^{2}\alpha_{z}}+\widehat{\kappa}\left(\alpha^{2}\alpha_{z}-1\right)=\frac{\widehat{F}_{\text{ext}}}{\alpha^{2}}\,,
\label{eq:implicit-alpha-alphaZ}
\end{equation}
implicitly determining the components $\alpha$, $\alpha_{z}$ of the elastic deformation gradient $\mathbb{A}$. The growth law \eqref{eq:growth-law-compressible-nH} takes the form 
\begin{equation}
\dot{\gamma}_{z}\gamma_{z}^{-1}=\widehat{\mu}^{*}-\widehat{W}+\alpha_{z}\widehat{F}_{\text{ext}}-2\widehat{\chi} \gamma_z\left(\frac{\gamma_z-1}{\left(\gamma_z+1\right)^{3}}\right)\, .
\label{eq:bar-growth-law}
\end{equation}
Here, the $zz$-component of the Eshelby stress is $\widehat{S}_{zz}=\widehat{W}-\alpha_z \widehat{F}_z$. 

Previously we have seen that depending on the chemical potential, i.e. the availability of nutrients, our model can exhibit three regimes: Collapse, size control and exponential growth. Now, we explore also how the growth penalty $\widehat{\chi}$ affects the regions. To this end, it is instructive to fully write out \eqref{eq:bar-growth-law} which means that we expand the free energy density $\widehat{W}$ according to \eqref{eq:quadratic-form-free-energy-nondim}. Then, \eqref{eq:bar-growth-law} can be written as 
\begin{equation}
\dot{\gamma}_{z}\gamma_{z}^{-1}=\widehat{\mu}^{*}+\widehat{\chi}h\left(\gamma_{z}\right)+\widehat{c}\label{eq:homogeneous-bar-short}
\end{equation}
where we introduced the shorthand 
\begin{align}
h\left(\gamma_{z}\right) & =-\frac{4}{(\gamma_{z}+1)^{3}}+\frac{4}{(\gamma_{z}+1)^{2}}-\frac{1}{2}\label{eq:h-bar-definition}\\
\widehat{c} & =-\frac{1}{2}\left(2\alpha^{2}+\alpha_{z}^{2}-3\right)-\frac{1}{2}\widehat{\kappa}\left(\alpha^{2}\alpha_{z}-1\right)^{2}+\log\left(\alpha^{2}\alpha_{z}\right)+\alpha_{z}\widehat{F}_{\text{ext}}\label{eq:c-bar-definition}
\end{align}
Note that all $\gamma_{z}$-dependence is contained in $h\left(\gamma_{z}\right)$. When the external force $\widehat{F}_{\text{ext}}$ is given, $\alpha$ and $\alpha_{z}$ can be determined via \eqref{eq:implicit-alpha-alphaZ}, and so $\widehat{c}$ according to \eqref{eq:c-bar-definition} is then just a constant.

The right-hand side of the dynamical law \eqref{eq:homogeneous-bar-short}, that is $\widehat{\mu}^{*}+\widehat{\chi}h\left(\left|\mathbb{G}\right|\right)+\widehat{c}$, has a relatively straightforward structure: Its lowest and highest values, respectively, are at $\gamma_{z}=0$ and $\gamma_{z}=1/2$. We can thus define the different types of dynamical behavior as follows: 
\begin{align}
\text{collapse}:\qquad & \widehat{\mu}^{*}+\widehat{\chi}h\left(1/2\right)+\widehat{c}<0\\
\text{size control}:\qquad & h\left(0\right)\leq h\left(\gamma_{z}\right)\leq h\left(1/2\right)\\
\text{unbounded growth}:\qquad & \widehat{\mu}^{*}+\widehat{\chi}h\left(0\right)+\widehat{c}>0\, .
\end{align}
The idea of defining these regions is centered around whether the right hand side of \eqref{eq:homogeneous-bar-short}, shown as a blue curve in the insets of Fig. \ref{fig:1D-model}B and C, crosses zero or not. In the collapse region, the blue curve is completely below zero, allowing no equilibrium of the dynamics, only shrinking. In the unbounded growth region, the curve is always above zero and once again no equilibrium is possible. In the size control region, the curve crosses zero at two points, one of which (the one with the greater value of $\gamma_z$) is the only stable equilibrium point, ensuring that the final size is always reached regardless of initial conditions.

This is precisely how the red (collapse), blue (size control) and yellow (unbounded) regions are defined in Fig. \ref{fig:1D-model}A. The boundary between the collapse and size control region is defined by $\widehat{\mu}^{*}+\widehat{\chi}h\left(1/2\right)+\widehat{c}=0$, and is thus a linear relationship between $\widehat{\mu}^{*}$ and $\widehat{\chi}$, forming the left line of the ``V''. The boundary between the size control region and the unbounded growth region is defined by $\widehat{\mu}^{*}+\widehat{\chi}h\left(0\right)+\widehat{c}=0$, forming the right line of the ``V''. 
In Fig. \ref{fig:1D-model}A, we can see that the scaled growth penalty parameter $\widehat{\chi}$ opens up the size control region in a "V"-shape: The sharp edge of the "V" corresponds to $\widehat{\chi}=0$, i.e. the classical case, in which size control is not present and you can only have either collapse or exponential growth. As $\widehat{\chi}>0$ increases, the size control region gets wider. The insets show example trajectories with different initial conditions for $\gamma_z$ at $\widehat{t}=0$. In the size control region, regardless of initial conditions, the same final size is reached, since there is only one stable equilibrium point $\hat{\dot{\gamma}}_z=0$. Figs.  \ref{fig:1D-model}B and C show slices through this diagram at constant values $\widehat{\chi}>0$ and $\widehat{\chi}=0$, respectively, contrasting the present and classical theory. In the classical case there is no size control regime, whereas in the present theory it "opens up" in the form of the blue region between collapse and unbounded growth.


\subsection{\label{sec:compressible-spheroid}3D: Compressible spheroid}

\begin{figure}
\hfill{}\includegraphics{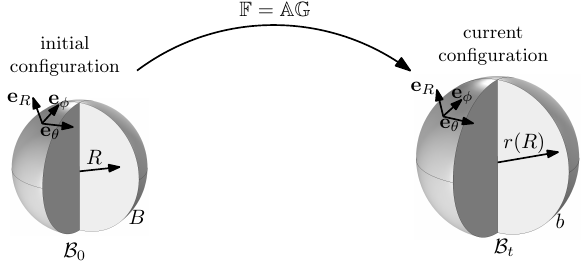}\hfill{}

\caption{\label{fig:relationships-sphere}Sketch of morphoelastic decomposition for a ball. In the initial configuration $\mathcal{B}_{0}$ the ball is unstressed and undeformed, corresponding to the pre-growth stage. It has a radius $R$ and its boundary is at $R=B$, i.e. $ R\leq B$. The growth tensor $\mathbb{G}$ instructs how vectors from $\mathcal{B}_{0}$ are mapped into the reference configuration, $\mathcal{B}_{r}$ (not sketched here), and the elastic deformation gradient $\mathbb{A}$ maps vectors from $\mathcal{B}_{r}$ to the current configuration, $\mathcal{B}_{t}$. In the current configuration, the boundary of the ball is at $r=b$, which corresponds to its actual size.}
\end{figure}

\subsection{Kinematics}

We consider the case of a compressible growing neo-Hookean spheroid (i.e. a ball of cells). We work in spherical polar coordinate basis $\left\{ \mathbf{e}_{R},\mathbf{e}_{\theta},\mathbf{e}_{\phi}\right\} $ (the same basis vectors apply to both initial and current configurations). The deformation map is given by $\mathbf{x}=r\left(R,t\right)\mathbf{e}_{R}$, where $r$ is the current radial coordinate and $R$ the initial radial coordinate (we will not explicitly refer to reference coordinates here). The deformation gradient tensor $\mathbb{F}$ with these geometric restrictions will have the structure $\mathbb{F}=\frac{\partial r}{\partial R}\mathbf{e}_{R}\otimes\mathbf{e}_{R}+\frac{r}{R}\left(\mathbf{e}_{\theta}\otimes\mathbf{e}_{\theta}+\mathbf{e}_{\phi}\otimes\mathbf{e}_{\phi}\right).$ We shall assume the same coordinate basis as in the previous expression for all relevant tensors in this geometry ($\mathbb{A}$, $\mathbb{G}$, $\mathbb{T}$, $\mathbb{S}$) and will from now on write their components in index notation. The elastic deformation gradient takes the form $\mathbb{A}=\text{diag}\left(\alpha_{R},\alpha_{\theta},\alpha_{\theta}\right)$, and the growth tensor is $\mathbb{G}=\text{diag}\left(\gamma_{R},\gamma_{\theta},\gamma_{\theta}\right)$. To summarize, in index notation (with the basis $\left\{ \mathbf{e}_{R},\mathbf{e}_{\theta},\mathbf{e}_{\phi}\right\} $ implied), we have 
\begin{equation}
\mathbb{F}=\text{diag}\left(\frac{\partial r}{\partial R},\frac{r}{R},\frac{r}{R}\right),\qquad\mathbb{A}=\text{diag}\left(\alpha_{R},\alpha_{\theta},\alpha_{\theta}\right),\qquad\mathbb{G}=\text{diag}\left(\gamma_{R},\gamma_{\theta},\gamma_{\theta}\right)\:.
\label{eq:disk-decomposition-1}
\end{equation}
In the initial configuration $\mathcal{B}_{0}$, the domain boundary is located $R=B$, see illustration in Fig. \ref{fig:relationships-sphere}.

\subsection{Mechanics}
Given that all deformations are diagonal in the coordinate basis considered here, the Cauchy stress is also diagonal $\mathbb{T}=T_{R}\mathbf{e}_{R}\otimes\mathbf{e}_{R}+T_{\theta}\left(\mathbf{e}_{\theta}\otimes\mathbf{e}_{\theta}+\mathbf{e}_{\phi}\otimes\mathbf{e}_{\phi}\right).$ In the present geometry the Cauchy stress tensor according to \eqref{eq:Cauchy-stress-compressible} has the following components:
\begin{equation}
T_{\left\{ R,\theta\right\} }=\frac{G}{\left|\mathbb{A}\right|}\left(\alpha_{\left\{ R,\theta\right\} }^{2}-1\right)+\kappa\left(\left|\mathbb{A}\right|-1\right)\,.
\label{eq:constitutive-law}
\end{equation}
Notice that $T_{\theta}-T_{R}=G\left(\alpha_{\theta}^{2}-\alpha_{R}^{2}\right)/|\mathbb{A}|$.  We assume that the  mechanical equilibrium holds $\nabla_{\mathbf{x}}\cdot\mathbb{T}=0$ in the presence of a hydrostatic pressure $\mathbb{T}=p_\text{ext}\mathds{1}$ at the boundary of the ball $\partial \mathcal{B}_{t}$, that is
\begin{equation}
\frac{\partial T_{R}}{\partial R}=\frac{2r'}{r}\left(T_{\theta}-T_{R}\right)=\frac{2Gr'}{r}\left(\frac{\alpha_{\theta}^{2}-\alpha_{R}^{2}}{\left|\mathbb{A}\right|}\right),\qquad T_{R}=p_{\text{ext}}\quad\text{at}\quad R=B\,.
\label{eq:stress-balance-continuous}
\end{equation}
Once $T_{R}$ is solved for, the hoop stress is computed as $T_{\theta}=T_{R}+G\left(\alpha_{\theta}^{2}-\alpha_{R}^{2}\right)/|\mathbb{A}|$. 

Next, we non-dimensionalize the system according to \eqref{eq:scaling}. In addition, we scale all lengths by the initial disk size $B$, that is $r=B\widehat{r}$ and $R=B\widehat{R}$, as well as scaling the external pressure by the shear modulus, that is $p_\text{ext}=G\, \widehat{p}_\text{ext}$ Together, the constitutive law \eqref{eq:constitutive-law} and the stress balance \eqref{eq:stress-balance-continuous} provide a boundary value problem which we refer to as the ``spatial problem'', as for fixed $\gamma_{R}$, $\gamma_{\theta}$ it is a system of coupled ODEs in space:
\begin{align}
\widehat{T}_{R} &=-\widehat{\kappa}+\left(\frac{\widehat{\kappa}\widehat{r}^{2}}{\widehat{R}^{2}\gamma_{\theta}^{2}\gamma_{R}}+\frac{\widehat{R}^{2}\gamma_{\theta}^{2}}{\widehat{r}^{2}\gamma_{R}}\right)\frac{\partial\widehat{r}}{\partial\widehat{R}}-\frac{\widehat{R}^{2}\gamma_{\theta}^{2}\gamma_{R}}{\widehat{r}^{2}(\partial\widehat{r}/\partial\widehat{R})}
, & \widehat{r}=0\quad\text{at}\quad\widehat{R}=0\label{eq:full-sphere-kinematic}\\
\frac{\partial\widehat{T}_{R}}{\partial\widehat{R}} & =\frac{2\left[\widehat{r}^{2}\gamma_{R}^{2}-\widehat{R}^{2}\gamma_{\theta}^{2}\left(\partial\widehat{r}/\partial\widehat{R}\right)^{2}\right]}{\widehat{r}^{3}\gamma_{R}}, & \widehat{T}_{R}=\widehat{p}_\text{ext}\quad\text{at}\quad\widehat{R}=1\label{eq:full-sphere-stress-balance}
\end{align}
with any imposed functions $\gamma_{R}(\widehat{R})$, $\gamma_{\theta}(\widehat{R})$. The non-dimensional hoop stress is obtained as $\widehat{T}_{\theta}=\widehat{T}_{R}+(\alpha_{\theta}^{2}-\alpha_{R}^{2})/|\mathbb{A}|$.

\begin{figure}[t]
	\hfill{}
	\includegraphics[width=0.75\textwidth]{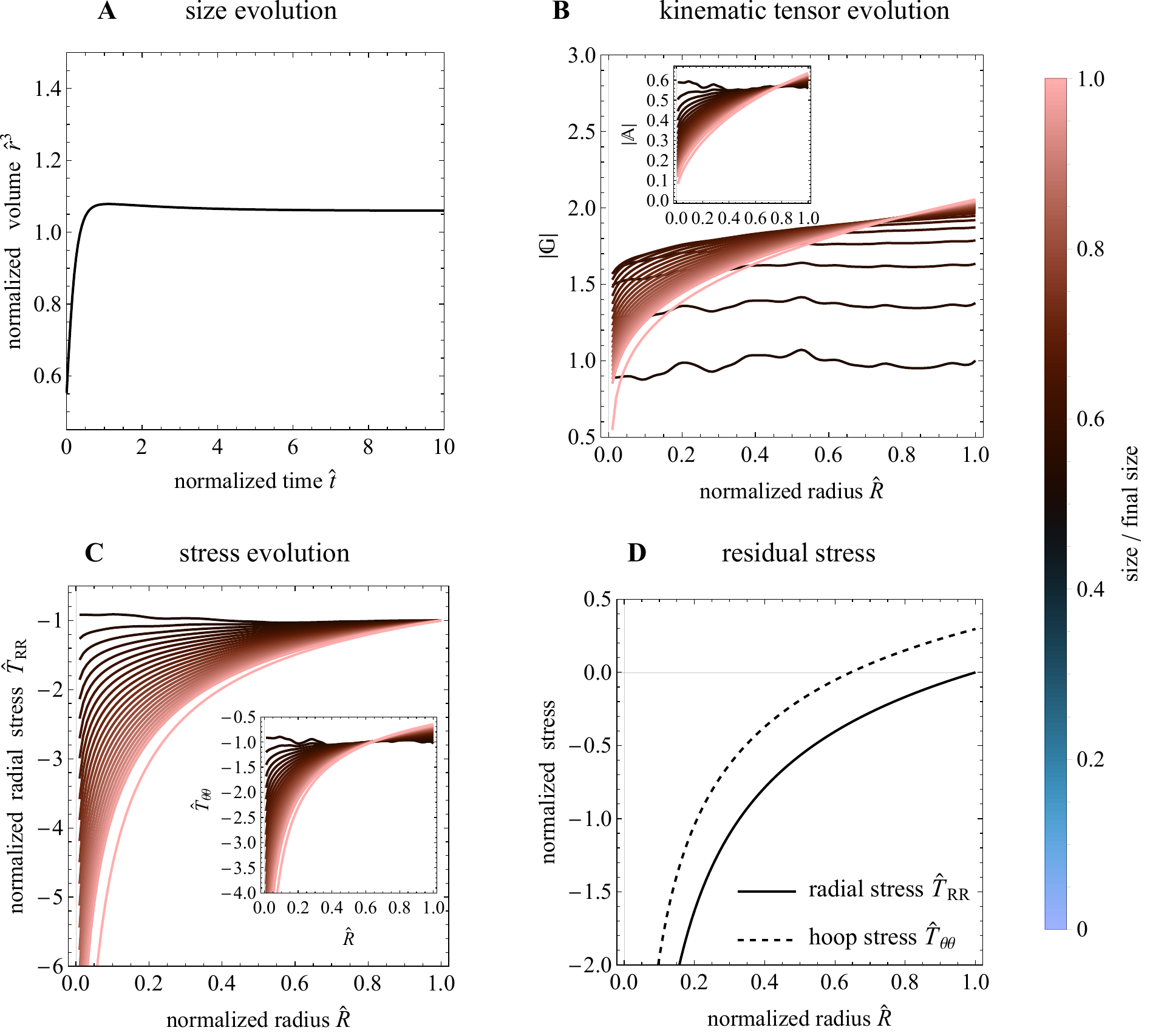}\hfill{}
	\caption{Qualitatively realistic scenario of multicellular spheroid growth. \textbf{A.} Evolution of the spheroid volume divided by the initial volume. \textbf{B.} Evolution of the determinants of the kinematic tensors $\mathbb{G}$ and $\mathbb{A}$. The initial conditions are noisy; the final (equilibrium) profile of  $|\mathbb{G}|$  is smooth.  \textbf{C.} Evolution of the stress profile. Due to the external compressive pressure on the spheroid, the final (equilibrium) stresses (rose color) are compressive. \textbf{D.} After the equilibrium state has been reached with the external pressure acting on the spheroid, the spheroid is removed from the experimental setup and returned to a no-pressure boundary condition $\widehat{p}_\text{ext}=0$. This reveals the residual stress in the system, with the hoop stress being tensile at the periphery and compressive at the center, which is consistent with cutting experiments of spheroids. Parameters are  $\widehat{\kappa}=1$, $\widehat{\chi}=6.5$,  $\widehat{\mu}^{*}=2$, $\widehat{\sigma}^{*}=0.15$, $\widehat{p}_\text{ext}=-1$.}
	\label{fig:realistic-scenario}
\end{figure}

\subsection{Growth}

We now reduce the growth law \eqref{eq:growth-law-compressible-nH} to a form appropriate for the present spherical geometry. Since the tensors $\mathbb{A}$ and $\mathbb{T}$ are diagonal and share the same basis, they commute. Therefore, the non-dimensional form of the Eshelby stress \eqref{eq:Eshelby-stress-definition} takes the  simplified form $\widehat{\mathbb{S}}=-\widehat{W}\mathds{21}+\left|\mathbb{A}\right|\widehat{\mathbb{T}}.$ Further, we decompose $\widehat{\mathbb{S}}^{*}$ into a hydrostatic part and a deviatoric part. In the component notation adopted above,
\begin{equation}
\widehat{\mathbb{S}}^{*}=\text{diag}\left(\widehat{\sigma}^{*},-\frac{1}{2}\widehat{\sigma}^{*},-\frac{1}{2}\widehat{\sigma}^{*}\right)+\text{diag}\left(\widehat{\mu}^{*},\widehat{\mu}^{*},\widehat{\mu}^{*}\right)\label{eq:deviatoric-hydrostatic-splot}
\end{equation}
where, as in previous notation, the hat in $\widehat{\sigma}^*$ and $\widehat{\mu}^*$ means that they are scaled by the shear modulus $G$. In this form, $\widehat{\mathbb{S}}^{*}$ is split into its deviatoric and hydrostatic part: the deviatoric part is $\text{dev }\widehat{\mathbb{S}}^{*}=\frac{1}{3}\text{tr }\widehat{\mathbb{S}}^{*}=\widehat{\mu}^{*}\mathds{1}$, and the hydrostatic part is $\text{hyd }\widehat{\mathbb{S}}^{*}=\widehat{\mathbb{S}}^{*}-\frac{1}{3}\text{tr }\widehat{\mathbb{S}}^{*}=\text{diag\ensuremath{\left(\widehat{\sigma}^{*},-\frac{1}{2}\widehat{\sigma}^{*},-\frac{1}{2}\widehat{\sigma}^{*}\right)}}$. Taking this into account, we write the non-dimensional growth law as
\begin{align}
\frac{\hat{\dot{\gamma}}_{R}}{\gamma_{R}} & =\widehat{\sigma}^{*}+\widehat{\mu}^{*}-\widehat{W}+\left|\mathbb{A}\right|\widehat{T}_{R}-2\widehat{\chi}\left|\mathbb{G}\right|\left(\frac{\left|\mathbb{G}\right|-1}{\left(\left|\mathbb{G}\right|+1\right)^{3}}\right) & \gamma_{R}=\gamma_{R}^{0}\quad\text{at}\quad\widehat{t}=0\label{eq:full-sphere-gamma-R}\\
\frac{\hat{\dot{\gamma}}_{\theta}}{\gamma_{\theta}} & =-\frac{1}{2}\widehat{\sigma}^{*}+\widehat{\mu}^{*}-\widehat{W}+\left|\mathbb{A}\right|\widehat{T}_{\theta}-2\widehat{\chi}\left|\mathbb{G}\right|\left(\frac{\left|\mathbb{G}\right|-1}{\left(\left|\mathbb{G}\right|+1\right)^{3}}\right) & \gamma_{\theta}=\gamma_{\theta}^{0}\quad\text{at}\quad\widehat{t}=0\, .\label{eq:full-sphere-gamma-Theta}
\end{align}
The energy $\widehat{W}$ is given by \eqref{eq:quadratic-form-free-energy-nondim}. The determinants of the kinematic tensors, in our geometry, are 
\begin{equation}
\left|\mathbb{A}\right|=\frac{\widehat{r}^{2}}{\widehat{R}^{2}}\left(\frac{\partial\widehat{r}}{\partial\widehat{R}}\right)\frac{1}{\gamma_{R}\gamma_{\theta}^{2}},\qquad\left|\mathbb{G}\right|=\gamma_{R}\gamma_{\theta}^{2}\,.
\end{equation}

In summary, for the whole boundary value problem we must simultaneously solve the constitutive law \eqref{eq:full-sphere-kinematic}, the stress balance \eqref{eq:full-sphere-stress-balance}, and the two equations for growth dynamics \eqref{eq:full-sphere-gamma-R}, \eqref{eq:full-sphere-gamma-Theta}. We describe our numerical approach to solving this system of partial algebraic-differential equations in \ref{sec:numerics}.

Fig. \ref{fig:realistic-scenario} shows a scenario that qualitatively mimics the situation in real spheroids under external pressure. Fig. \ref{fig:realistic-scenario}A shows a particular scenario with size control. The simulation starts with smooth noisy randomly generated initial conditions for $\gamma_R^0$, $\gamma_\theta^0$. Fig. \ref{fig:realistic-scenario}B shows the evolution of $|\mathbb{G}|$ and $|\mathbb{A}|$, which start out noisy. Eventually, they each converge to smooth profiles. Fig. \ref{fig:realistic-scenario}C shows the evolution of the stress profiles (Cauchy stress in  radial and hoop  directions, $T_R$ and $T_\theta$, respectively). The stress profile starts out as nearly uniform and nearly isotropic, $\widehat{\mathbb{T}}\approx \widehat{p}_\text{ext}\mathds{1}$, deviating from uniformity only slightly due to the noisy initial profile. It evolves to a smooth, non-uniform and anisotropic profile. Due to the external pressure, which is compressive, the stress profiles are all compressive, consistent with the results of  \cite{alessandri2013cellular} where external pressure is achieved via a polymer coating of spheroids.  Finally, Fig. \ref{fig:realistic-scenario}D  shows the equilibrium configuration of Fig. \ref{fig:realistic-scenario}C, but with the external pressure removed. This setup reveals the residual stress which has been built due to non-uniform anisotropic growth. The hoop stress is tensile at the periphery, and compressive at the center, in line with the experiments of \cite{stylianopoulos2012causes} where qualitatively the same stress pattern was observed in spheroids through the opening of radially cut slices of spheroids \cite{guillaume2019characterization,colin2018experimental}.  In summary, the discussion of Fig. \ref{fig:realistic-scenario} reveals how the deviatoric part of the homeostatic stress, controlled by the parameter $\widehat{\sigma}^*$, induces a spatial inhomogeneity in the profiles of stress and kinematic tensors and causes the buildup of residual stress in the spheroid.   

\begin{figure}[t]
\hfill{}
\includegraphics[width=0.7\textwidth]{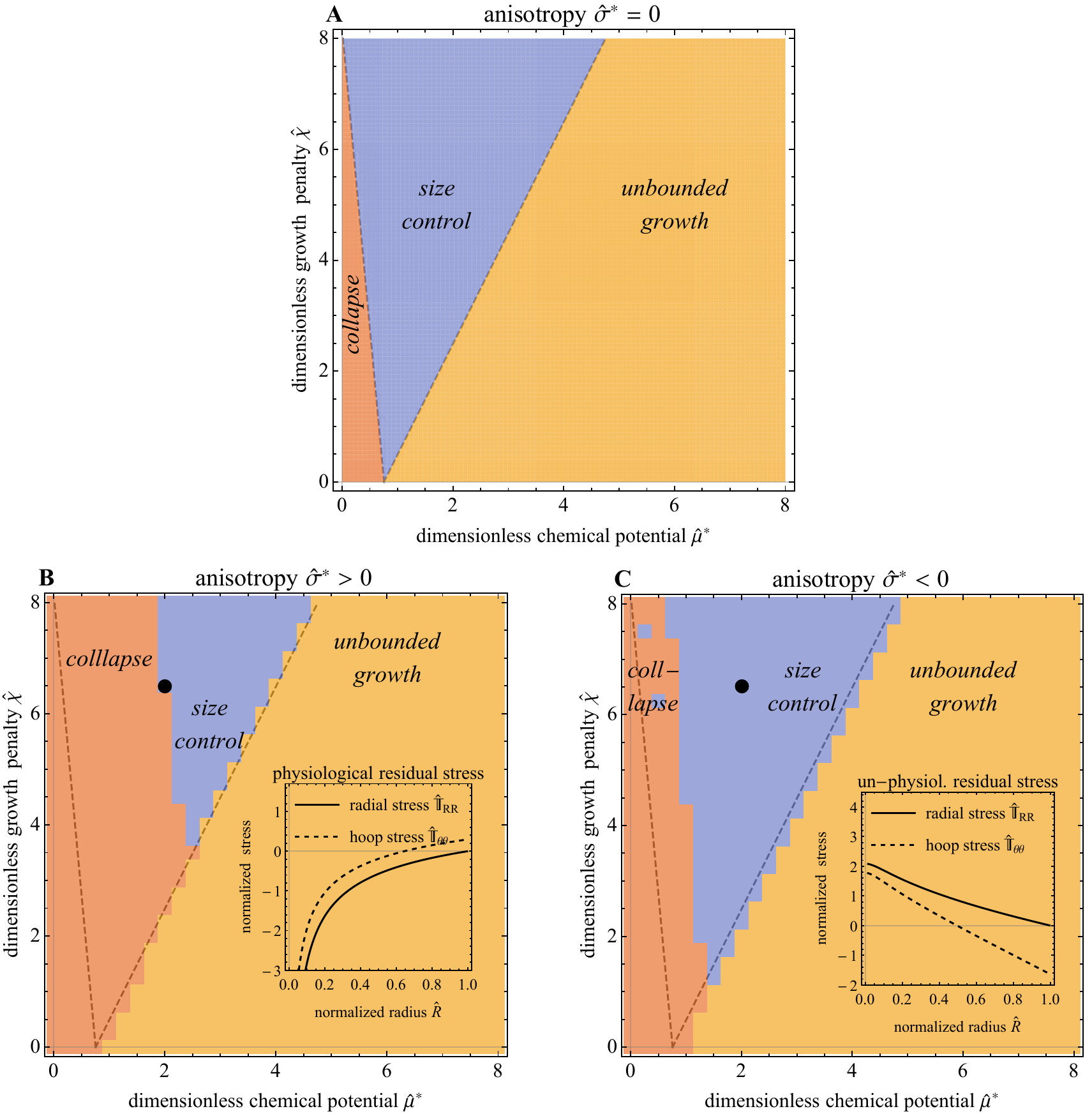}\hfill{}
\caption{ How the anisotropy of the homeostatic stress shifts the size control boundary. \textbf{A.} When the homeostatic stress tensor is isotropic $\widehat{\mathbb{S}}^*=\widehat{\mu}^* \mathds{21}$, we obtain a "V"-like separation between the region of collapse, size control and exponential growth, qualitatively identical to Fig. \ref{fig:1D-model}.  \textbf{B.} When the homeostatic stress tensor is anisotropic in favor of the radial component ($\widehat{\sigma}^*>0$), the size control boundary shrinks compared to the scenario where homeostatic stress is isotropic ($\widehat{\sigma}^*=0$, dashed line). This causes a realistic residual stress profile (inset), with tensile hoop stress at the boundary, which is consistent with experiments in which the spheroid is cut and opens. \textbf{C.} In the opposite scenario, where then anisotropy of the homeostatic stress tensor is in favor of the hoop component ($\widehat{\sigma}^*<0$), the size control boundary also  shrinks compared to the scenario where homeostatic stress is isotropic ($\widehat{\sigma}^*=0$, dashed line). This causes compressive hoop stress at the boundary (inset), which is not consistent with experiments.  Parameters are  $\widehat{\kappa}=1$, $\widehat{p}_\text{ext}=-1$. The black dots in both subfigures highlight the parameters $\widehat{\mu}^*=2$, $\widehat{\chi}=6.5$. For subfigures B and C, the deviatoric part of the homeostatic stress is $\widehat{\sigma}^*=0.15$ and $\widehat{\sigma}^*=-0.4$, respectively. } 
\label{fig:effect-of-sigma}
\end{figure}

\section{\label{sec:residual-stress-and-size-regulation}Residual stress and size regulation}

When the homeostatic stress tensor has no deviatoric part ($\widehat{\sigma}^*=0$), the deformation field in the spheroid is spatially homogeneous, i.e. the stress in the spheroid is isotropic and the same everywhere, matching the externally imposed pressure for all times. This case is illustrated in a \ref{fig:effect-of-sigma}A, in which the collapse region (red) and the unbounded growth region (yellow) are separated by a "V"-shaped size control region (blue). The region boundaries for this case $\widehat{\sigma}^*=0$ are worked out semi-analytically in \ref{sec:homogeneous}. This result is qualitatively identical to the uniaxially growing bar which was explored in Fig. \ref{fig:1D-model}, in which deformations and stresses inside the bar were also completely uniform. In both cases, the bar and the spheroid at $\widehat{\sigma}^*=0$, once the external force or pressure is removed, the stress in the system goes back to zero, the systems are free of residual stress. 

As was shown in Fig. \ref{fig:realistic-scenario}B and C, when the homeostatic stress has a non-zero deviatoric part ($\widehat{\sigma}^*\neq 0$), quantities like the Cauchy stress tensor $\mathbb{T}$ and the kinematic tensors $\mathbb{F}$, $\mathbb{A}$ and $\mathbb{G}$ become spatially non-uniform and anisotropic, even if they start in a completely uniform and isotropic state. As Fig. \ref{fig:realistic-scenario}D demonstrates, when the external pressure is removed from the spheroid, the Cauchy stress does not vanish, which means that the growth dynamics has built up residual stress due to the fact that the homeostatic stress tensor is anisotropic ($\widehat{\sigma}^*\neq 0$).

In Fig.  \ref{fig:effect-of-sigma}B and C, we explore how the deviatoric part of the homeostatic stress tensor $\widehat{\sigma}^*$ shifts the regions of collapse, size control and unbounded growth compared to the "V"-shaped case $\widehat{\sigma}^*=0$ shown in Fig.  \ref{fig:effect-of-sigma}A. Indeed, the region boundaries of \ref{fig:effect-of-sigma}A are shown in Fig. \ref{fig:effect-of-sigma}B and C as the gray dashed lines for comparison. We see that in both cases $\widehat{\sigma}^*>0$ (\ref{fig:effect-of-sigma}B) and  $\widehat{\sigma}^*<0$ (\ref{fig:effect-of-sigma}C), the size control region (blue) shrinks in comparison to the case where the homeostatic stress is isotropic $\widehat{\mathbb{S}}^*=\widehat{\mu}^* \mathds{21}$ (which corresponds to  $\widehat{\sigma}^*=0$). 

The direct comparison of the cases $\widehat{\sigma}^*>0$ and  $\widehat{\sigma}^*<0$ allows us to see which of these cases is consistent with experimentally observed results of residual stress in spheroids. In \cite{stylianopoulos2012causes}, it was observed that cutting tumor spheroids with a knife leads to an opening of the "lips" at the cut. This is consistent with a tensile hoop stress at the boundary \cite{stylianopoulos2012causes,colin2018experimental,guillaume2019characterization}, but would be inconsistent with a compressive hoop stress, since in that case the spheroid would stay closed \cite{goriely2017mathematics}.  This allows us to identify \ref{fig:effect-of-sigma}B as a physically plausible scenario, in which the residual hoop stress is consistent with experiments, and in which a finite size is reached (Fig. \ref{fig:realistic-scenario}A), which is consistent with experimental observations \cite{alessandri2013cellular}.

\begin{figure}[t]
	\hfill{}
	\includegraphics[width=0.8\textwidth]{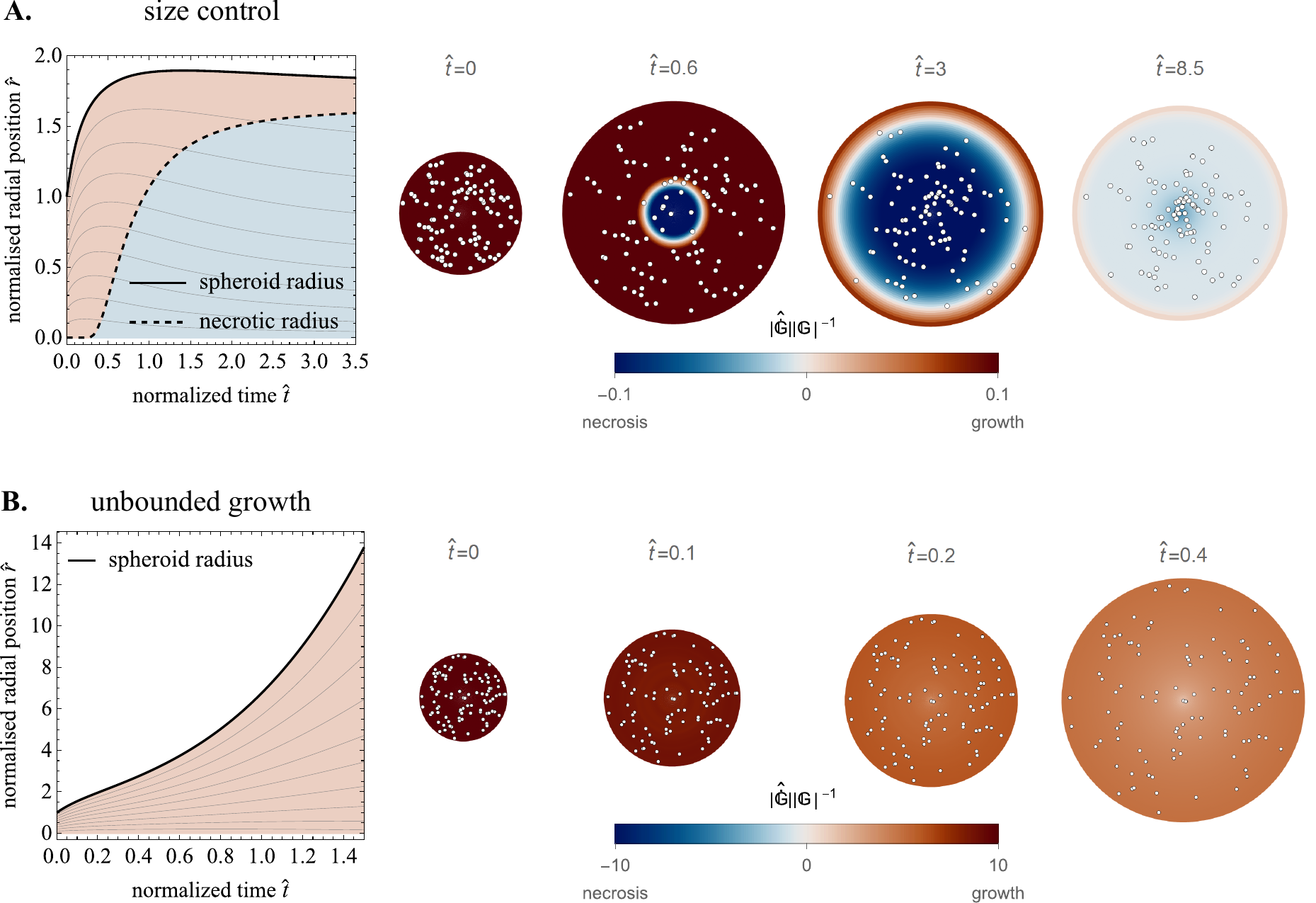}\hfill{}
	\caption{ Transient scenarios and the necrotic core. \textbf{A.} The evolution of the spheroid radius (black, solid), alongside the paths of internal material points (gray, solid). Initially, all material points move away from the core of the spheroid in a rapid growth phase, but afterwards there is some backflow of material towards the necrotic core. The necrotic radius (dashed) separates the necrotic core $|\hat{\dot{\mathbb{G}}}|<0$ (blue shaded region) from a proliferating region $|\hat{\dot{\mathbb{G}}}|>0$ (light red shaded region). The spheroid at initial time $\widehat{t}=0$ is marked with 100 randomly placed points. After the initial growth phase, the necrotic core (white ring) becomes visible, coinciding with the flow of marked points towards the center. Around $\widehat{t}=8.5$, the dynamics is close to equilibrium ($|\hat{\dot{\mathbb{G}}}|$ is close to zero), as material points have clearly moved to the center compared to $\widehat{t}=0$. Parameters are , $\widehat{\mu}^*=4.50$, $\widehat{\sigma}^*=0.49$. 
			\textbf{B.} In the case of unbounded growth, there is no backflow of material and no necrotic core. The marked material points expand almost isometrically.  Parameters are: $\widehat{\mu}^*=6.5$, $\widehat{\sigma}^*=0.52$. The following parameters were used for both simulations A. and B.:  $\widehat{p}_\text{ext}=0$, $\widehat{\kappa}=1$, $\widehat{\chi}=10$. }
	\label{fig:necrotic-core}
\end{figure}

\section{\label{sec:necrotic-core}Transient dynamics and the necrotic core}

The presence of a necrotic core in the center of a growing spheroid is generally attributed to a nutrient depletion within the core \cite{tracqui2009biophysical}. The classical explanation is that avascular tumor growth relies on the delivery of oxygen and nutrients from surrounding host tissue and their consumption by tumor cells. The nutrient availability affects cell proliferation and death, leading to the formation of a concentric structure with a necrotic center, a dormant middle layer, and a highly active outer layer, due to the sharp differences in chemical concentration. In mechanosensitive tumor models, the necrotic core usually emerges due to a coupling of oxygen diffusion and stress feedback, see e.g. \cite{walker2022towards,xue2016biochemomechanical}.  

Perhaps surprisingly, the present model suggests an alternative explanation for the formation of the necrotic core. In the presence of anisotropic homeostatic stress, a transient phase can emerge in which a necrotic core becomes apparent, as long as the spheroid is in the size control regime, see Fig. \ref{fig:necrotic-core}A. There, we visualize the transient initial growth phase which occurs before eventually all growth stops ($\hat{\dot{\mathbb{G}}}=\mathbf{0}$ for all $\widehat{R}$). This transient phase can be broadly subdivided into two phases: Initially (between $\widehat{t}=0$ and $\widehat{t}=0.3$ in Fig. \ref{fig:necrotic-core}A), the tumor grows rapidly without a necrotic core. After that, but before the spheroid reaches its largest size, a necrotic core emerges (the necrotic core is defined as those radial points for which $|\hat{\dot{\mathbb{G}}}|<0$). The effect of the necrotic can be seen clearly when observing randomly marked points on a spheroid: There is a material flow towards the center as the points are "attracted" into the necrotic core. Just before the spheroid reaches a steady state in which any material flow stops $\hat{\dot{\mathbb{G}}}=\mathbf{0}$, the initially marked points have clearly moved towards the center. This observation is qualitatively consistent with experiments in which cells of freely growing spheroids have been stained and showed the same trend of outward followed by inward flow (see Fig. 4d in \cite{delarue2013mechanical}). In the case of an unbounded freely growing spheroid, on the other hand, no necrotic core emerges as material points expand nearly isometrically, leading to unbounded proliferation of the tissue.

\section{\label{sec:Discussion}Discussion}

The key idea in this article is that we make a modification of the free energy which we identify with the physical effect of size regulation. The modification is led by the insight that for a cell to grow, or shrink, there is an energetic cost. \chg{Some active processes that consume chemical energy from ATP hydrolysis are directly involved in the control of the local tissue swelling such as the cells ion pumping and the process of endocytosis. This energy consumption can be maintained thanks to the cell metabolic system that consumes nutrients  \cite{yang2021physical}. Our model do not explicitly consider these detailed processes and their coupling with the extracellular matrix as it operates only with the variable $\mathbb{G}$ characterizing the swelling at the tissue scale. It is only broadly consistent with the fact that these swelling processes consume energy, which should be reflected in the free energy of the system. Our way to fix this energetic cost is to phenomenologically select a dependence that is consistent with the expected behavior of a cell colony even in the absence of mechanical interactions between the cells. }

\begin{table}[b]
	\centering{\resizebox{0.85\textwidth}{!}{%
			\begin{tabular}{>{\centering}m{5cm}>{\centering}m{2cm}>{\centering}m{2cm}>{\centering}m{2cm}>{\centering}m{2cm}}
				\toprule 
				& Shi-Lei et al. \cite{xue2016biochemomechanical} & Ambrosi et al. \cite{ambrosi2017solid} & Walker et al. \cite{walker2022towards} & present\tabularnewline
				\midrule 
				residual stress as in experiments & $\times$ & $\checkmark$ & only at periphery & $\checkmark$\tabularnewline
				\midrule 
				necrotic core is present & $\checkmark$ & $\times$ & $\checkmark$ & $\checkmark$\tabularnewline
				\midrule 
				oxygen diffusion is modeled & $\checkmark$ & $\checkmark$ & $\checkmark$ & $\times$\tabularnewline
				\midrule 
				derived from thermodynamics & $\checkmark$ & $\times$ & $\times$ & $\checkmark$\tabularnewline
				\bottomrule
	\end{tabular}}}
	
	\caption{\label{tab:comparison-of-models}Comparison of tumor models. The category "residual stress as in experiments" refers to tensile hoop stress at the periphery and compressive hoop stress near the core, and compressive radial stress throughout, which would lead to an opening of the spheroid upon incision \cite{stylianopoulos2012causes,ambrosi2017mechanobiology,colin2018experimental,guillaume2019characterization}. In \cite{walker2022towards}, this appears to be true only very closely to the edge of the spheroid, while in the bulk the profile is reversed.}
\end{table}

\chg{\subsection{Comparison of tumor models}}
A number of experimental observations about growing multicellular spheroids are well established, and capturing them all in a single model has not been achieved previously. While the poroelastic model of Xue and co-workers \cite{xue2016biochemomechanical} captures the growth profiles observed in experiments (necrosis towards center, high proliferation at the rim), their stress distribution is not fully consistent with experiments, as the hoop stress in their model is compressive. This would not cause an opening of the spheroid when it is cut, as reported in \cite{stylianopoulos2012causes,colin2018experimental,guillaume2019characterization}. On the other hand, in the model of Ambrosi and co-workers \cite{ambrosi2017solid}, the stress distribution is captured, with tensile hoop stress at the rim which is consistent with experiments. The same model, however, explicitly does not include the possibility of a necrotic core.  Finally, the model of Walker and co-workers \cite{walker2022towards} qualitatively captures the right trends  for the residual stress of a spheroid, although only close to the boundary (in the bulk, the stress profiles reverse). The same model also captures a necrotic core. However, this model relies on a non-local feedback mechanism, i.e. any cell must have access to the largest magnitude compressive stress experienced by all cells in the entire spheroid.

The present model simultaneously provides a qualitatively correct residual stress profile (Fig. \ref{fig:realistic-scenario}D) and captures the inward flow of material associated with necrosis (Fig. \ref{fig:necrotic-core}A). All this is achieved through a local feedback mechanism derived from a thermodynamical framework, requiring only a single phase material (i.e. no poroelasticity) and no model of oxygen diffusion. A caveat is that the necrotic core emerges only in a transient phase and eventually vanishes as growth stops, rather than being a true "dynamic equilibrium" state. A comparison of the features of the different models discussed here is shown in Table \ref{tab:comparison-of-models}.

\begin{figure}
	\hfill{}
	\includegraphics[width=1.0\textwidth]{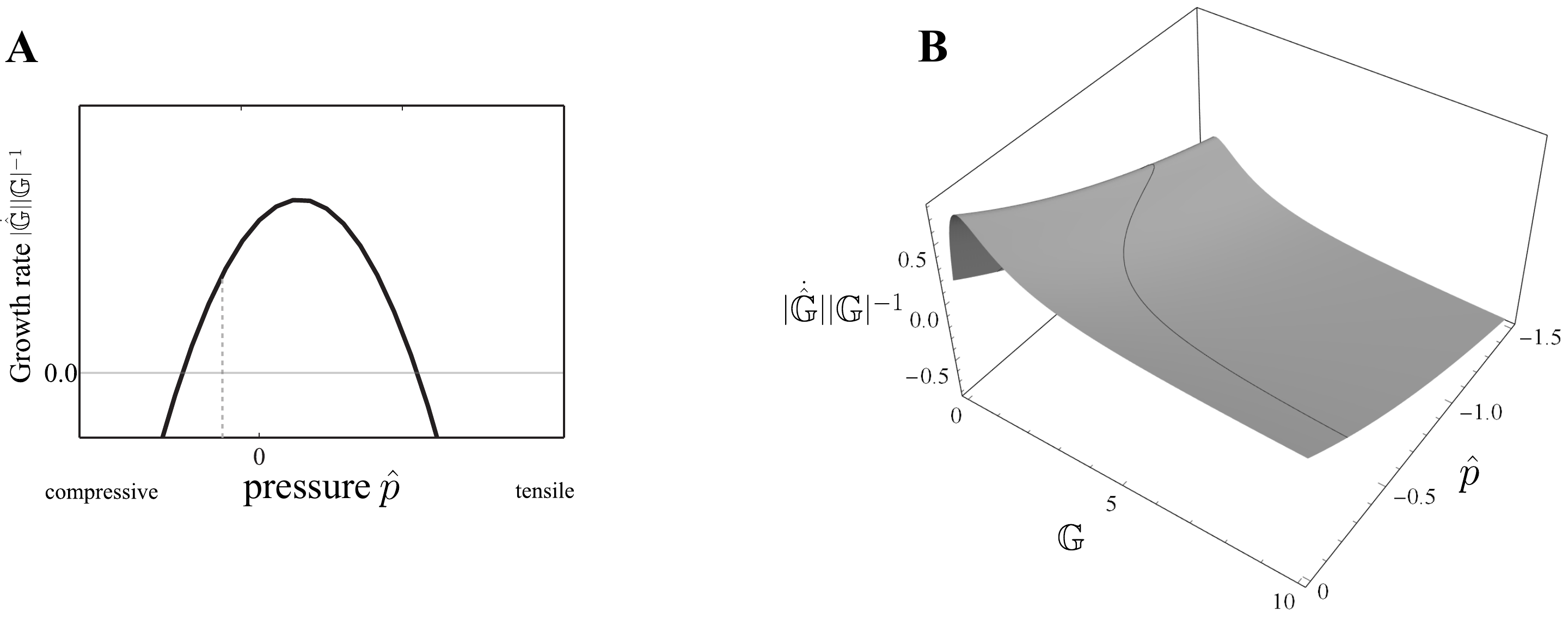}\hfill{}
	\caption{\chg{Mechanical feedback in Shraiman's models compared to  the present model. \textbf{A.} Sketch of Shraiman's non-monotonic mechanical feedback mechanism, adapted after \cite{hufnagel2007mechanism} Fig. 3. Too much tension or too much compression leads to resorption of the tissue, while the optimal growth rate is achieved when the tissue is in tension. Mechanical feedback is seen as an essential ingredient to size control in \cite{hufnagel2007mechanism}. \textbf{B.} In the present model, even in the absence of mechanical feedback ($\widehat{p}=0$), a finite size can be reached, and the carrying capacity of the cell colony is set by the chemical potential $\widehat{\mu}^*$. Mechanical feedback reduces the size of the spheroid in the presence of increasing compression (black contour). Parameters are: $\widehat{\mu}^{*}=0.7$, $\widehat{\chi}=1$, $\widehat{\kappa}=1$.}}
	\label{fig:shraiman-vs-us}
\end{figure}

\chg{\subsection{Different approaches to size regulation in the presence of mechanical feedback }}

An influential model \chg{of a feedback mechanism between growth and stress was presented by Boris Shraiman in \cite{shraiman2005mechanical}, and was later extended to produce size regulation in \cite{hufnagel2007mechanism}. }The mechanical feedback in both these works exhibits a non-monotonic growth rate (see \cite{shraiman2005mechanical} Fig. 3 and \cite{hufnagel2007mechanism} Fig. 3), which we reproduced in Fig. \ref{fig:shraiman-vs-us}A. Shraiman postulates a growth-mechanical coupling according to which a bit of tension is optimal for growth, but too much tension kills the tissue. This feedback leads to unbounded exponential growth \cite{shraiman2005mechanical}. \chg{In the perspective of morphoelasticity, a similar feedback can be obtained if one takes the classical law \eqref{eq:generic-growth-law} and makes the kinetic coefficient $\mathbb{K}$ dependent on stress. Similarly to \cite{shraiman2005mechanical}, this type of feedback produces unbounded growth. In order to add the possibility of attaining an asymptotic size, Shraiman and co-authors \cite{hufnagel2007mechanism} combined the mechanical feedback mechanism shown in Fig. \ref{fig:shraiman-vs-us}A with a diffusing chemical species (a morphogen) as well as a thresholding mechanism by which growth stops if the morphogen level falls below a certain threshold. In this point of view, mechanical feedback is essential for growth arrest.

A fundamental difference in our model is that even in the absence of mechanics, growth arrest is possible if the chemical potential $\widehat{\mu}^*$ lies in the size control regime. The case of no mechanics can be seen in Fig. \ref{fig:no-mechanical-feedback} as a non-monotonic relationship between the growth rate and $|\mathbb{G}|$ which is a proxy for the system size. Fig. \ref{fig:shraiman-vs-us}B shows how this non-mechanical scenario, captured by the $\widehat{p}=0$ plane, is modified when the pressure $\widehat{p}$ becomes increasingly compressive: The growth rate is shifted downwards, and the equilibrium size (represented by the black contour) decreases, which is consistent with the expectation that a more compressed spheroid will reach a smaller size. 

A similar point of view is taken by Ambrosi and co-authors in \cite{ambrosi2017solid}. There, in the absence of mechanical feedback ($c=1$ and $\text{tr }S=0$ in their notation), the growth of a spheroid follows a logistic (Verhulst) growth curve. Mechanical feedback in their model is activated through a prescribed non-uniform concentration profile decaying exponentially from the oxygen-saturated boundary.  A contribution of the present work is to give an energetic and thermodynamical interpretation to a growth profile of logistic type, and to relate the carrying capacity of the logistic growth to the external chemical potential $\mathbb{S}^*$. In our view, the anisotropy of the potential tensor $\mathbb{S}^*$ activates mechanical feedback which also affects the carrying capacity of the spheroid. \\}  

\chg{\subsection{Size regulation in mixture models}}

\chg{In the present work, we focus on size control in the classical kinematic decomposition framework in which the growing biological material, i.e. its cells, their constituents, and the extracellular matrix, are all treated as a single field. To consider the interplay between multiple constituents, such as solid cell components and water in a multicellular spheroid, or elastin and collagen in arteries, the concept of a mixture has been proposed 
there have been several mixture theories in which size regulation has been demonstrated, at the cost of adding considerable complexity to the modeling. These can be broadly split in two categories: 
\begin{enumerate}
    \item Solid-fluid mixtures, of which the most canonical one incorporates one solid and one fluid phase, are modeled by poroelasticity. In this view, the solid cell components and extracellular matrix are interpenetrated by a fluid. The classical framework of poroelasticity \cite{coussy2004poromechanics} has been adapted to growth theories with a multiplicative decomposition \cite{fraldi2018cells}. Using a similar approach, a poroelastic model of a growing spheroid  was able to produce growth arrest \cite{xue2016biochemomechanical} (this model and its caveats are contrasted in Table \ref{tab:comparison-of-models}). However, due to the relative complexity of such models, and an almost complete lack of analytical solutions, an explicit understanding of the minimal ingredients needed for size regulations is missing. 
    \item Solid-solid mixtures model the fact that soft tissues are often composed of different load-bearing constituents such as cells and their collagen matrix. Constrained mixture theory simplifies the problem by assuming equal velocities for all solid phases \cite{humphrey2002constrained,cyron2017growth,ambrosi2019growth}. This theory is particularly relevant for studying growth and remodeling in tissues where different constituents have different turnover dynamics. A drawback is that the history of the dynamics of each individual phase must be tracked and evaluated in ``hereditary integrals'' which requires considerable computational effort. To investigate questions such as dynamical stability and the final size of an artery in healthy conditions vs. conditions in which aneurysms appear, the constrained mixture theory approach has been extended to a dynamical perspective known as ``mechanobiological stability"''\cite{cyron2014vascular,cyron2014mechanobiological,latorre2019mechanobiological}. In addition to the usual constitutive law describing stress-strain relations, mechanobiological stability approaches require additional constitutive laws for each constituent in form of a mass production and mass removal rate. Paired with a suitable homogenization method to eliminate the aforementioned hereditary integrals, a mechanobiological stability analysis based on dynamical systems theory has shown that artery models produce a stable organ size via a homeostasis mechanism resembling \eqref{eq:generic-growth-law}, and proposed explanations for aneurysms via a mechanobiological instability \cite{cyron2014vascular,cyron2014mechanobiological,latorre2019mechanobiological}. In this case, a homeostatic feedback can produce a finite size and stable equilibrium due to the force balance between the different phases. But if only one phase is present, this is not possible in general. This is seen in the 1D bar example in Fig. \ref{fig:1D-model}C. However, if an extra  bar is added in parallel, a finite size for the composite system becomes possible \cite{erlich2020role}.  Analogously to the two competing bars in parallel balancing forces, in the case of multicellular spheroids, the cells and the extracellular matrix (ECM) \cite{dolega2020mechanical} represent two phases that balance forces with each other. Furthermore, the growth dynamics of the two phases is interlinked: The two processes of cell swelling during its growth and the ECM swelling are coupled since cells are able to deposit their ECM as well as digest it. 
\end{enumerate}

One may speculate that mixture approaches may in the future provide mechanisms that underlie the phenomenological form of $W_\text{g}$ proposed in this article. Indeed, in the case of poroelasticity, the free energy traditionally includes not only the elastic contribution of the solid phase, but also a contribution from the coupling of solid and fluid phases. Since the fluid phase is linked to $|\mathbb{G}|$ via mass conservation relationships, a poroelastic approach resulting in a free energy of the form $W=W_\text{el}(\mathbb{A})+W_\text{g}(|\mathbb{G}|)$ is conceivable.  It is also possible that after an appropriate homogenization of the hereditary integrals of constrained mixture theory, a free energy  of the type $W_\text{g}$ could emerge due to the differential growth of the cells and ECM  phases.}

\chg{\subsection{Model limitations}}

The presence of the new growth term in the free energy density establishes size control, meaning that if the parameters of the system are contained in the blue region in Fig. \ref{fig:effect-of-sigma}, the system will reach the same size independently of initial conditions. This is contrary to the classical system \eqref{eq:generic-growth-law}, in which for some prescribed value of the externally imposed chemical potential $\mathbb{S}^*$ (also known as the homeostatic Eshelby stress tensor) the system will either collapse or grow without bounds , see Fig. \ref{fig:1D-model}C. The present model endows the classical model with size control (Fig. \ref{fig:1D-model}B). In the present model, in its simplest form, biochemistry is reduced to an externally imposed constant scalar chemical potential  $\mu^*$, bringing a simplicity and clarity that allows the investigation of the relationships between biochemistry, mechanics, residual stress and size (Fig. \ref{fig:effect-of-sigma}B and C and Fig. \ref{fig:1D-model}B). 

\chg{Our model is a simplification that considers only one cell type and one resource availability. This allows us to illustrate the coupling between mechanics and logistic growth, but it does not account for other important aspects of tumor biology. For example, our model does not distinguish between apoptosis and necrosis, but only captures the loss of cell mass. Moreover, our model does not include the phenotypic heterogeneity of cancer cells that depends on the oxygen and nutrient levels in different regions of the tumor. These factors could be incorporated in our model by integrating a diffusive concentration field into the kinetic coefficient $\mathbb{K}$ similarly to \cite{ambrosi2007growth}, through a spatiotemporal dependence of $\widehat{\mu}^*$, and/or by adding internal variables in the free energy. However, such extensions are beyond the scope of our work and can be found in other models, such as \cite{villa2021modeling,fraldi2018cells}.\\}

\chg{\subsection{An energetic penalty for growth $W_\text{g}$ in the context of physics and biology} }
Most mechanical theory of growth do not take into account an energetic cost for growth $W_\text{g}$, which includes single phase theories \cite{dicarlo2002growth,epstein2000thermomechanics,ambrosi2008stress}, poroelastic growth theories \cite{gao2016embryo,fraldi2018cells}, or chemo-mechanical theories \cite{ambrosi2007growth,ciarletta2012mass}.  
While the term $W_\text{g}$ in \eqref{eq:quadratic-form-free-energy} is often absent from growth theories, when viewed from plasticity theory, the presence of an anelastic field in the free energy is not a surprise. As discussed in Section \ref{sec:analogy-strain-gradient-plasticity}, in plasticity theory, the free energy is typically additively decomposed into a contribution due to elasticity $W_\text{el}$ and a defect energy $W_\text{d}$ which is the energy stored in the crystal lattice due to dislocations: $W=W_\text{el}+W_\text{d}$. 

We argue that in growth theories, an analog of the defect energy $W_{\text{d}}$ should also be included to penalize a \emph{competition} between cells and their environment. This energy is the growth penalty $W_\text{g}$. \chg{For a cell colony on a petri dish, as discussed above, this competition is mainly due to nutrient availability. However, in different contexts, when more details are known about the mechanics and microstructure, $W_\text{g}$ may need to be adapted to the biological setting.  }  In an epithelial layer, this may be the competition between cortical tension and contractility of cells \cite{farhadifar2007influence}. In a setup of multiple adjacent layers, the competition may be the differential growth between a cell layer and adjacent extracellular matrix \cite{oltean2016tissue,harmansa2022growth}. If the tissue is best described as a solid-fluid mixture (capturing, for instance, the solid skeleton of the cell and intracellular fluid), the competition may be between the solid phase and the fluid phase \cite{gao2016embryo}. Finally, if the tissue is best described by a solid-solid mixture (capturing collagen, elastin and smooth muscle cells in an artery) \cite{gleason20042,alford2008growth}, the competition may be between the solid phases of the mixture. \\

\chg{
\subsection{Towards a microstructural theory of cellular growth through water mobility and ion fluxes}
}
In the present situation, unlike in plasticity, an established microstructural theory rigorously linked to a coarse-grained free energy does not exist yet. We thus propose a choice of $W_{\text{g}}$ that divides the possible outcomes into three scenarios: The death of a cell colony, its unbounded proliferation, and a controlled state where a certain size of the cell colony is maintained. A future challenge will be to justify the form of $W_{\text{g}}$ based on microscopic laws of cell growth. Solid cell components, such as cell organelles or the cytoskeleton, only occupy about 20\% of the total cell volume \cite{cadart2019physics}. Therefore, \chg{a crucial} problem of understanding how cells grow in volume, and consequently how the tissue grows in volume, is to understand how water flows in and out of cells. Water mobility in and out of cells relies on the permeation of water through the plasma membrane, which can be regulated by aquaporin channels, which are permeable to water but not ions \cite{kedem1958thermodynamic}, as well as ion pumps which actively consume energy. 

In this view, water mobility and ion fluxes take a highly important role in the regulation of growth, and should therefore be linked to the coarse grained field $W_\text{g}(|\mathbb{G}|)$ . While historically in growth models of living tissue, fluid phases were not considered \cite{ambrosi2004role,taber2002theoretical,taber2009towards,goktepe2010multiscale,Bowden2015wound,erlich2018mechanical,almet2021role}, there has recently been considerable interest in combining elements from morphoelasticity and poroelasticity \cite{xue2016biochemomechanical,fraldi2018cells,ambrosi2017solid} or discrete vertex models explicitly tracking fluxes between cells \cite{cheddadi2019coupling}. In parallel, there are new developments coupling the electrochemistry of ion fluxes, mechanics of cell volume regulation, active pumping through ATP hydrolysis, all expressed in a close to equilibrium thermodynamical framework respecting the Onsager relationships \cite{cadart2019physics,deshpande2021chemo,dolega2021extra,duclut2019fluid}. These developments could be stepping stones towards a full microscopic theory that would provide the underlying mechanisms that lead to the coarse-grained form  $W_\text{g}(|\mathbb{G}|)$.\\

\chg{
\subsection{Possible experiments in spheroids and \textit{Drosophila} towards understanding size}
}
An excellent way of testing predictions of the growth law \eqref{eq:growth-law-non-dim} is through experiments in  multicellular spheroids. These lab-made cell cultures have clear spherical symmetry, can be tested with high throughput methods \cite{alessandri2013cellular}, and established protocols for cutting experiments in order to quantify the residual stress in the system \cite{stylianopoulos2012causes,ambrosi2017solid}. Furthermore, the mechanical environment of multicellular spheroids can be controlled very precisely. For instance,  multicellular spheroids can be compressed by supplementing the culture medium with large Dextran molecules that cannot permeate the spheroid pores. The imposed osmotic pressure leads to an interstitial fluid flow that dehydrates the spheroid leading to its compression \cite{delarue2013mechanical,delarue2014compressive,dolega2017cell}. These experimental tools can allow us to look at the long timescale behavior ($\sim$ one week) of multicellular spheroids and reveal their final size as a function of tightly controlled external pressure ($\widehat{p}_\text{ext}$ in our model) or the chemical microenvironment ($\widehat{\mu}^*$). In particular, our model predicts that different initial sizes would lead to the same final size in the size control regime. Whether spheroids of different initial sizes would reach the same final size is still an unanswered question, and testing this hypothesis would be an important  first step towards confirming the growth law  \eqref{eq:growth-law-non-dim}.  

While multicellular spheroids are artificial \textit{in vitro} systems, a good living biological model system for some of the same questions of growth termination is the \textit{Drosophila} wing disc due to its relative simplicity. Attempts to understand its size termination have  spawned a number of chemical and chemo-mechanical models (\cite{hufnagel2007mechanism,aegerter2007model,wartlick2011dynamics,averbukh2014scaling,aguilar2018critical}) which however fail to give a coherent picture of the relationship between size, diffusion of morphogens, residual stress and growth. Recent studies reveal that the geometry of this system is a multi-layered sandwich-like structure \cite{harmansa2022growth,nematbakhsh2020epithelial} which is clearly residually stressed, as shown by a cutting experiment \cite{harmansa2022growth}. Further experimental progress on the \textit{Drosophila} wing disc, in particular the measurement of asymptotic size of the wing disc, may make it feasible in the future to test the predictions of our model in this developing organ. \\ \, \\ \,  

Ultimately, the question of size regulation is complex and will likely involve many mechanisms including biochemical signaling, morphogen diffusion, oxygen diffusion and, in its absence, cell death (\cite{dolega2020mechanical}), as well as water fluxes between cells and osmotic regulation (\cite{cheddadi2019coupling}). Still, a formulation of feedback laws that respect the dissipation inequality, and that have sound microscopic underpinnings, is a physically grounded approach which  may lead to an understanding of the mechanisms that control the buildup of stress and control of size across different model systems. Such an endeavor is of relevance to the developmental biology community \cite{aegerter2007model,aegerter2012integrating}, the physics community \cite{aguilar2018critical,hufnagel2007mechanism}, and to the mechanics community \cite{ambrosi2015active,pettinati2016finite}. \\

\section*{Computational codes}

The numerical implementation was done in Mathematica 12.1. The notebooks generating all the figures can be downloaded at: \url{https://github.com/airlich/size-control/}.

\section*{Acknowledgments}
 We thank Lev Truskinovsky, Larry Taber and Davide Ambrosi for insightful individual discussions on growth models, as well as each for their detailed feedback on our manuscript. We also thank Giuseppe Zurlo for in-depth discussions on growth and size regulation.  

\appendix

\section{\label{sec:Explicit-form-of-dfdA-dfdG}Explicit form of $\partial f/\partial\mathbb{A}$ and $\partial f/\partial |\mathbb{G}|$}

In order to evaluate the dissipation inequality \eqref{eq:Dissipation-over-Jg}, we must first evaluate the expressions $\partial f/\partial\mathbb{A}$ and $\partial f/\partial\left|\mathbb{G}\right|$. To this end, the following identities are very useful, see Matrix Cookbook (\cite{petersen2008matrix}) equations (49), (108) and (52), respectively: $\frac{\partial I_{1}}{\partial\mathbb{A}}=2\mathbb{A}$, $\frac{\partial I_{3}}{\partial\mathbb{A}}=2I_{3}\mathbb{A}^{-\mathsf{T}}$ and $\frac{\partial\left|\mathbb{A}\right|}{\partial\mathbb{A}}=\left|\mathbb{A}\right|\mathbb{A}^{-\mathsf{T}}$. 

Further, we define the trace and double contraction operator in the following ways: With some arbitrary second order tensors $\mathbb{U}$ and $\mathbb{M}$, we have $\mathbb{U}:\mathbb{M}=\text{tr}(\mathbb{U}^{\mathsf{T}}\mathbb{M})=\text{tr}(\mathbb{U}\mathbb{M}^{\mathsf{T}})=U_{ij}M_{ij}$ as well as $\mathds{21}:\mathbb{U}=\text{tr}(\mathbb{U})$ . 

To calculate $\partial f/\partial\mathbb{A}$ for the dissipation inequality, we refer to the free energy in the reference configuration \eqref{eq:free-energy-in-reference-config}, and utilize the chain rule to get $\partial f/\partial\mathbb{A}=\left(\partial W/\partial\mathbb{A}\right)/\rho_{r}$. Using the form \eqref{eq:quadratic-form-free-energy} for the free energy, we get
\begin{align}
\frac{\partial f}{\partial\mathbb{A}}= & \frac{\left|\mathbb{A}\right|}{\rho_{r}}\left\{ \frac{G}{\left|\mathbb{A}\right|}\left(\mathbb{A}-\mathbb{A}^{-\mathsf{T}}\right)+\kappa\left(\left|\mathbb{A}\right|-1\right)\mathbb{A}^{-\mathsf{T}}\right\}  
\label{eq:dfdA}\\
\frac{\partial f}{\partial\left|\mathbb{G}\right|}= & 2\frac{\chi}{\rho_{r}}\frac{\left|\mathbb{G}\right|-1}{\left(\left|\mathbb{G}\right|+1\right)^{3}}
\label{eq:dfdG}
\end{align}

With these expressions, the Cauchy stress \eqref{eq:Cauchy-stress-compressible} and the growth law \eqref{eq:growth-law-compressible-nH} can be obtained.

\section{\label{sec:homogeneous}The role of mechanical feedback in size regulation}

In this section, we derive semi-analytically the boundaries of the diagram \ref{fig:effect-of-sigma}A, corresponding to the case  $\widehat{\sigma}^*=0$  for a growing spheroid. The approach we take is practically identical to the uniaxially growing bar described in subsection \ref{sec:1d-bar}. The basic assumption here is that in the spheroid, all deformations and stresses are spatially uniform and isotropic for all times. In this case, all relevant tensor quantities ($\mathbb{A},\mathbb{G},\widehat{\mathbb{T}},\widehat{\mathbb{S}},\widehat{\mathbb{S}}^{*}$) are isotropic and spatially uniform. 

Now, we make the following assumption for homogeneous deformations:
\begin{equation}
\mathbb{G}=\left|\mathbb{G}\right|^\frac{1}{3}\mathds{1},\qquad\mathbb{A}=\left|\mathbb{A}\right|^\frac{1}{3}\mathds{1},\qquad \widehat{r}=\left(\left|\mathbb{A}\right|\left|\mathbb{G}\right|\right)^{\frac{1}{3}}\widehat{R}\,.\label{eq:homogeneous-compressible-sphere}
\end{equation}
Here, it is assumed that both $\left|\mathbb{G}\right|$ and $\left|\mathbb{A}\right|$ are spatially uniform. Next,  we insert the assumption \eqref{eq:homogeneous-compressible-sphere} into the definition of the Cauchy stress tensor \eqref{eq:Cauchy-stress-compressible}, 
\begin{equation}
\widehat{p}_{\text{ext}}=\frac{\left(\left|\mathbb{A}\right|^{2/3}-1\right)}{\left|\mathbb{A}\right|}+\widehat{\kappa}(\left|\mathbb{A}\right|-1)\,.
\label{eq:detA-positive-root}
\end{equation}
The positive root of this equation provides $\left|\mathbb{A}\right|$ as a function of $\widehat{p}_{\text{ext}}$ and $\widehat{\kappa}$. Next, we return to the growth law \eqref{eq:full-sphere-gamma-R}, \eqref{eq:full-sphere-gamma-Theta}. In addition to the present assumption, we also impose $\widehat{\sigma}^*=0$, which means that the homeostatic stress tensor is isotropic, $\widehat{\mathbb{S}}^*=\widehat{\mu}^* \mathds{21}$. The growth law then becomes isotropic and can be written as 
\begin{equation}
\frac{\hat{\dot{\left|\mathbb{G}\right|}}}{\left|\mathbb{G}\right|}=\widehat{\mu}^{*}-\widehat{W}+\left|\mathbb{A}\right|\widehat{p}_{\text{ext}}-2\widehat{\chi}\left|\mathbb{G}\right|\left(\frac{\left|\mathbb{G}\right|-1}{\left(\left|\mathbb{G}\right|+1\right)^{3}}\right) \,.\label{eq:yes-MF-reduced}
\end{equation}
Here, the scaled Eshelby stress is isotropic, $\widehat{\mathbb{S}}=\left(\widehat{W}-\left|\mathbb{A}\right|\widehat{p}_{\text{ext}}\right)\mathds{1}$. 

Note that when we expand \eqref{eq:yes-MF-reduced}, we can write it as 
\begin{equation}
	\frac{\hat{\dot{\left|\mathbb{G}\right|}}}{\mathbb{G}}=\widehat{\mu}^{*}+\widehat{\chi}h\left(\left|\mathbb{G}\right|\right)+\widehat{c}\; ,\label{eq:homogeneous-dynamics-short}
\end{equation}
where we introduced the shorthand
\begin{align}
	h\left(\left|\mathbb{G}\right|\right) & =-\frac{4}{(\left|\mathbb{G}\right|+1)^{3}}+\frac{4}{(\left|\mathbb{G}\right|+1)^{2}}-\frac{1}{2}\;,\label{eq:h-definition}\\
	\widehat{c} & =\frac{1}{2}\left(-3\left|\mathbb{A}\right|^{2/3}+2\log(\left|\mathbb{A}\right|)+3\right)-\frac{1}{2}\widehat{\kappa}(\left|\mathbb{A}\right|-1)^{2}+\left|\mathbb{A}\right|\widehat{p}_{\text{ext}}\; . \label{eq:c-definition}
\end{align}
Note that all $\left|\mathbb{G}\right|$-dependence is contained in $h\left(\left|\mathbb{G}\right|\right)$. When the external pressure $\widehat{p}_{\text{ext}}$ is given, $\left|\mathbb{A}\right|$ can be determined via \eqref{eq:detA-positive-root}, and so $\widehat{c}$ according to (\ref{eq:c-definition}) is then just a constant. 

The right hand side of the dynamical law (\ref{eq:homogeneous-dynamics-short}), that is $\widehat{\mu}^{*}+\widehat{\chi}h\left(\left|\mathbb{G}\right|\right)+\widehat{c}$, has a relatively straightforward structure: Its lowest and highest values, respectively, are at $\left|\mathbb{G}\right|=0$ and $\left|\mathbb{G}\right|=1/2$. We can thus define the different types of dynamical behavior as follows:
\begin{align}
	\text{collapse}:\qquad & \widehat{\mu}^{*}+\widehat{\chi}h\left(1/2\right)+\widehat{c}<0\\
	\text{size control}:\qquad & h\left(0\right)\leq h\left(\left|\mathbb{G}\right|\right)\leq h\left(1/2\right)\\
	\text{unbounded growth}:\qquad & \widehat{\mu}^{*}+\widehat{\chi}h\left(0\right)+\widehat{c}>0
\end{align}
This is precisely how the red (collapse), blue (size control) and yellow (unbounded) regions were defined in Fig. \ref{fig:effect-of-sigma}A. The boundary between the collapse and size control region is defined by $\widehat{\mu}^{*}+\widehat{\chi}h\left(1/2\right)+\widehat{c}=0$, and is thus a linear relationship between $\widehat{\mu}^{*}$ and $\widehat{\chi}$, forming the left line of the ``V''. The boundary between the size control region and the unbounded growth region is defined by $\widehat{\mu}^{*}+\widehat{\chi}h\left(0\right)+\widehat{c}=0$, the linear relationship between $\widehat{\mu}^{*}$ and $\widehat{\chi}$ thus form the right line of the ``V''.

\section{\label{sec:numerics}Numerical approach}

Here, we discuss our numerical approach to solving the dynamical growth of the compressible neo-Hookean spheroid \eqref{eq:full-sphere-gamma-R}, \eqref{eq:full-sphere-gamma-Theta}. This system of highly non-linear coupled partial differential equations  must be solved for $\gamma_{R}\left(R,t\right)$, $\gamma_{\theta}\left(R,t\right)$, given a set of initial conditions $\gamma_{R}^{0}\left(R\right)$, $\gamma_{\theta}^{0}\left(R\right)$ at $\widehat{t}=0$. We denote all initial conditions with a superscript zero. 

We approach the numerical solution via a method of lines discretization, in which the spatial variable $R$ is discretized with a finite difference scheme, whereas the temporal discretization is left to a standard method for the numerical solution of systems of non-linear ordinary differential equations such as fourth-order Runge Kutta. 

We discretize the spatial domain into $N+1$ node points, such that node $i=0$ corresponds to the disk center and $i=N-1$ corresponds to the disk periphery, whereas the last remaining node $i=N$ is a ghost node to accommodate spatial derivatives.

We denote the spatial discretization with a subscript index $i$, such that the continuum variable $\widehat{r}(\widehat{R},\widehat{t})$ in its spatially discretized version becomes $\widehat{r}_{i}(\widehat{t})$, and $\widehat{T}_{R}$$(\widehat{R},\widehat{t})$ becomes $\widehat{T}_{R,i}(\widehat{t})$. The components of the growth tensor become $\gamma_{R,i}(\widehat{t})$ and $\gamma_{\theta,i}(\widehat{t})$. The initial radial coordinate $\widehat{R}$ becomes $\widehat{R}_{i}=i\Delta\widehat{r}$ where $\Delta\widehat{r}=1/\left(N-1\right)$. To discretize all equations for our problem, we use a backward difference scheme for spatial derivatives:
\begin{equation}
\frac{\partial\widehat{r}}{\partial\widehat{R}}\rightarrow\frac{\widehat{r}_{i}(\widehat{t})-\widehat{r}_{i-1}(\widehat{t})}{\Delta\widehat{r}},\qquad\frac{\partial\widehat{T}_{R}}{\partial\widehat{R}}\rightarrow\frac{\widehat{T}_{R,i}(\widehat{t})-\widehat{T}_{R,i-1}(\widehat{t})}{\Delta\widehat{r}}\,.
\end{equation}

First, we focus on the discretization of the spatial problem, \eqref{eq:full-sphere-kinematic} and \eqref{eq:full-sphere-stress-balance}. The boundary conditions for the spatial problem are $\widehat{r}=0$ at $\widehat{R}=0$ and $\widehat{T}_{R}=0$ at $\widehat{R}=1$, the discrete versions become 
\begin{equation}
\widehat{r}_{0}(\widehat{t})=0\qquad\text{and}\qquad\widehat{T}_{R,N-1}(\widehat{t})=0\qquad\text{for all }\widehat{t}\:.
\end{equation}
Thus, the discretization of the spatial problem \eqref{eq:full-sphere-kinematic}, \eqref{eq:full-sphere-stress-balance} provides $2\left(N-1\right)$ algebraic relationships for the $2\left(N-1\right)$ unknowns $\widehat{r}_{1}(\widehat{t}),\ldots,\widehat{r}_{N-1}(\widehat{t})$ and $\widehat{T}_{R,0}(\widehat{t}),\ldots,\widehat{T}_{R,N-2}(\widehat{t})$. 

Next, we turn our attention to the full dynamical problem \eqref{eq:full-sphere-gamma-R}, \eqref{eq:full-sphere-gamma-Theta}. Discretized, these coupled PDEs become $2\left(N-1\right)$ coupled ordinary differential equations in time, which must be solved for the $2\left(N-1\right)$ unknowns $\gamma_{R,1}(\widehat{t}),\ldots,\gamma_{R,N-1}(\widehat{t})$ and $\gamma_{\theta,1}(\widehat{t}),\ldots,\gamma_{\theta,N-1}(\widehat{t})$. In summary, we have a differential algebraic system of $2\left(N-1\right)$ differential equations and an additional $2\left(N-1\right)$ algebraic constraints. 

To properly parameterize a differential-algebraic solver, we need initial conditions for all unknowns in the system, of which there are $4\left(N-1\right)$. We consider an initial state of the system in which the growth tensor is homogeneous (see Section ...), i.e. there is no residual stress in the system. The $2\left(N-1\right)$ initial conditions for the growth tensor are an isotropic, uniform field $g_{\text{init}}$:
\begin{equation}
\gamma_{R,1}^{0}=\ldots=\gamma_{R,N-1}^{0}=g_{\text{init}},\qquad\gamma_{\theta,1}^{0}=\ldots=\gamma_{\theta,N-1}^{0}=g_{\text{init}}\,.\label{eq:initial-conditions-G}
\end{equation}
For $\widehat{r}$ and $\widehat{T}_{R}$, the initial conditions must satisfy \eqref{eq:detA-positive-root}. With a prescribed external pressure $\widehat{p}_{\text{ext}}$, the initial condition for the Cauchy stress is $\widehat{T}_{R}=\widehat{p}_{\text{ext}}$ on the whole spatial domain, and $\widehat{r}=\left(\left|\mathbb{A}\right|\left|\mathbb{G}\right|\right)^{\frac{1}{3}}\widehat{R}$ on the whole domain. Here $\left|\mathbb{G}\right|$ is provided by the initial condition (\ref{eq:initial-conditions-G}), whereas the initial $\left|\mathbb{A}\right|$ can be found by calculating the positive root of \eqref{eq:detA-positive-root}, of which we denote the solution $\left|\mathbb{A}\right|=a_{\text{init}}^{3}$. Thus, the $2\left(N-1\right)$ initial conditions for the spatial problem are 
\begin{equation}
\widehat{r}_{i}^{0}=a_{\text{init}}g_{\text{init}}i\Delta\widehat{r}\qquad\text{for}\qquad i=1,\ldots,N-1\qquad\text{and}\qquad\widehat{T}_{R,i}^{0}=\widehat{p}_{\text{ext}}\qquad\text{for}\qquad i=0,\ldots,N-2\,.
\end{equation}
We use the \emph{Mathematica }function \texttt{NDSolve{[}{]}} to solve the differential-algebraic problem described here. In pseudocode, the full problem can be solved by 
\begin{equation}
\texttt{NDSolve[\{ODEs,AEs,ICs\},\{vars\},\{t,0,tend\}]}\:,\label{eq:NDSolve}
\end{equation}
where $\texttt{ODEs}$ is a shorthand for the $2\left(N-1\right)$ differential equations, $\texttt{AEs}$ for the $2\left(N-1\right)$ algebraic constraints, and $\texttt{ICs}$ for the $4\left(N-1\right)$ initial conditions, and $\texttt{tend}$ is a number for the final time of integration. The vector $\texttt{vars}$ of dimension $4\left(N-1\right)$ contains all variables to be solved for, that is 
\begin{equation}
\texttt{vars}=\left(\widehat{r}_{1},\ldots\widehat{r}_{N-1},\widehat{T}_{R,0},\ldots,\widehat{T}_{R,N-2},\gamma_{R,1},\ldots,\gamma_{R,N-1},\gamma_{\theta,1},\ldots,\gamma_{\theta,N-1}\right)^{\mathsf{T}}\,.
\end{equation}
The numerical approach (\ref{eq:NDSolve}) was used for Fig. \ref{fig:realistic-scenario}, Fig. \ref{fig:effect-of-sigma}B and C, and Fig. \ref{fig:necrotic-core}. The full code is provided in the repository \url{https://github.com/airlich/size-control/}.

\section*{\textemdash \textemdash \textemdash \textemdash \textemdash \textendash{}}

\bibliographystyle{elsarticle-harv}

\end{document}